\magnification=1000
\hsize =32true pc\vsize = 48true pc
\hoffset=.375 true in
\voffset=.5true in
\font\mybig=cmbx12 at 12pt

\font\mysmall=cmr8 at 8 pt
\font\eightit=cmti8

\input epsf

\def\c{\bf C}

\def\r{\bf R}

\def\hh{\cal H}

\def\ff{\cal F}

\def\ss{\cal S}

\def\ab{\cal A}

\def\qed{\diamondsuit}
\def\picture #1 by #2 (#3){
		\vbox to #2{
		\hrule width #1 height 0pt depth 0pt
		\vfill
		\special {picture #3}}}
\font\teneufm eufm10 
\font\seveneufm eufm7 
\font\fiveeufm eufm5
\newfam\eufm
\textfont\eufm\teneufm
\scriptfont\eufm\seveneufm
\scriptscriptfont\eufm\fiveeufm
\def\frak#1{{\fam\eufm\teneufm#1}}
\font\mybig=cmbx12 at 12pt 
\def\picture #1 by #2 (#3){
		\vbox to #2{
		\hrule width #1 height 0pt depth 0pt
		\vfill
		\special {picture #3}}}
\vskip 0.5 true in\centerline
{\bf \mybig Airy Functions for Compact Lie Groups.\/}
\vskip 0.5 true in
		\centerline {\bf Rahul N. Fernandez
 and V. S. Varadarajan \/}
		\bigskip
		\centerline {\it Department of Mathematics}
		\centerline{
		University of California, Los Angeles, CA 90095-1555, USA. \/}
\bigskip\noindent 
\bigskip\noindent
{{\bf Abstract.}  \mysmall  The classical Airy function has been generalised by Kontsevich to  a function of a matrix argument, which is an integral over the space of (skew) hermitian matrices of a unitary-invariant exponential kernel.   In this paper, the Kontsevich integral is generalised to integrals over the Lie algebra of an arbitrary connected compact Lie group, using exponential kernels invariant under the group.  The (real) polynomial defining this kernel is said to have the Airy property if the integral defines a function of moderate growth. A general sufficient criterion for a polynomial  to have the Airy property is given.  It is shown that an invariant polynomial on the  Lie algebra has the Airy property if its restriction to a Cartan subalgebra has the Airy property.  This result is used to evaluate these invariant integrals completely and explicitly on the hermitian matrices, obtaining formulae that contain those of Kontsevich as special cases.}
\bigskip\noindent
{\bf 1. Introduction.\/} The Airy function was discovered by the mathematician and astronomer {\it Sir George Biddell Airy\/}, who first introduced and discussed it in his paper [A] of  1838.  He defined it as the (improper Riemann) integral
$$
\int _0^\infty \cos \left ({\pi\over 2}\left ({\omega^3\over 3}-m\omega\right )\right )d\omega 
$$
and regarded it as a function of a real variable $m$.  Airy tabulated its values in his paper, and suspected that it could be expressed using known integrals;  it was later found to be expressible in terms of the Bessel function.  Numerous applications have since been found, both  in mathematics and in the physical sciences;  the reader is refered to [VS] for an encylop\ae dic survey of these.

A complex-valued version of Airy's original function, which we refer to as the {\it Airy function} or {\it Airy integral}, is given by
$$
A(x)=\int _{-\infty}^{\infty} e^{i((1/3)y^{3} -xy)}dy\qquad (x\in {\r}).
$$
Since this definition is classical,   it is appropriate to consider this integral as an improper Riemann integral, but one could equally well regard it as an integral over an appropriately chosen contour in the complex plane (see [VS], p.124).  It is a classical result that the Airy function satisfies the ordinary differential equation
$$
A''(x)+xA(x)=0.
$$
 It  can be shown that, up to a multiplying constant, it is the only solution to this equation having polynomial growth in $x$, and that it extends to an entire function on ${\c}$. One can also view it (this will be our point of view) as a distribution, being the Fourier transform of the tempered distribution $e^{iy^3/3}$.
\medskip
In [K], Kontsevich introduced the {\it matrix Airy function\/} $A(X)$,
$$
A(X)=\int _{{\hh}(n)} e^{i{\rm tr}({Y^{3}\over 3}-XY)}dY\qquad (X\in {\hh}(n)),
$$
where ${\hh}(n)$ is the vector space of $n\times n$ hermitian matrices.  This integral does not exist in the sense of Lebesgue, but it is well-defined as a distribution. The formal similarity of the matrix Airy function to the Airy function accounts for its name.  We sometimes refer to $A(X)$ as the {\it matrix Airy integral}. Kontsevich observes that $A(X)$ satisfies the elliptic partial differential equation
$$
\Delta A(X)+{\rm tr}(X)A(X) =0,
$$
and so is a smooth function by a standard theorem.  In his paper, he gives a moduli-theoretic interpretation of his function, and uses it to prove a conjecture of Witten [W].
\medskip
 The classical Airy function $A(x)$ is a special case of functions of the form
 $$
 A_p(x)=\int_{-\infty}^\infty e^{ip(y)-ixy}dy\qquad (x\in {\r}),
 $$
 where $p$ is a real polynomial on ${\r}$; the specialization $p(y)={y^3/3}$ gives $A(x)$. The matrix Airy function $A(X)$ is a special case of functions of the form
 $$
 A_{p, n}(X)=\int_{{\hh}(n)}e^{ip(Y)-i{\rm tr}(XY)}dY\qquad (X\in {\hh}(n)),
 $$
 where $p$ is a real ${\rm U}(n)$-invariant polynomial on ${\hh}(n)$; the specialization $p(Y)={\rm tr}(Y^3/3)$ gives the Kontsevich integral $A(X)$. Note that the definition of $A(X)$ can equally well be taken on the space of skew-hermitian matrices by changing $Y$ to $(-1)^{1/2}Y$, where the skew-hermitian matrices are interpreted as the Lie algebra of the unitary group ${\rm U}(n)$. 
 \medskip
 These remarks suggests a far-reaching generalization. Let $G$ be a connected compact Lie group and $V$ be a real finite-dimensional euclidean space, with scalar product $({\cdot} , {\cdot}),$ on which $G$ acts orthogonally. If $p$ is any $G$-invariant polynomial on $V,$ we formally define
 $$
 A_p(x)=\int_Ve^{ip(y)-i(x,y)}dy\qquad (x\in V).
 $$
 If $V={\hh}(n),\, G={\rm U}(n)$ and the $G$-action is given by $X\longmapsto gXg^{-1},$ we get $A_p(X)$. In view of Kontsevich's pioneering work, we refer to $A_p(X)$ as {\it Airy functions/integrals\/}. These definitions can be generalized even further by replacing ${\r}$ by a local or  finite field.  
\medskip
In this paper, we study these generalized Airy integrals for a specific finite dimensional module for an arbitrary connected compact Lie group, namely the adjoint representation. In sections 2 and 3 we develop their analytic theory, in particular obtaining sufficient conditions for an Airy integral to be a function of moderate growth that extends to an entire function. In Section 4 we establish the principle that an Airy integral on the Lie algebra of a connected compact Lie group is an entire function of moderate growth if the  result is true for its restriction to a Cartan subalgebra. This principle is used, in  Section 5, to evaluate some of the Airy integrals on the Lie algebra of ${\rm U}(n)$. Our formulae contain, as a special case, the formula obtained by Kontsevich in [K]. We hope that these generalizations of the Kontsevich integrals  will also have interesting moduli-theoretic interpretations. Finally, in the Appendix, we give a brief exposition, which we believe may be useful for people interested in working in this area, of Harish-Chandra's work on invariant differential operators and Fourier transforms on the Lie algebra of a compact Lie group. We use [V] as a general reference on semisimple Lie algebras and Lie groups. A significant part of the results of this paper forms the content of the thesis [F].
\bigskip\noindent
{\bf 2. The basic definitions.\/} Let $V$ be a real finite-dimensional euclidean space  with scalar product $({\cdot},{\cdot}),$ and let ${\ss}(V)$ be the Schwartz space of $V$ with its usual topology. The topological dual ${\ss}(V)^\prime$ of ${\ss}(V)$ is the space of tempered distributions on $V$. A function on $V$ is said to be of {\it polynomial growth\/} if it is  majorized by ${\rm const\/}(1+||v||)^m$ for some $m>0,$ where $||\cdot||$ is some norm on $V$, and of {\it moderate growth\/} if it is smooth and its derivatives (including itself) are of polynomial growth.  If $F$ is a measurable function $V\longrightarrow {\c}$ of polynomial growth, the map 
 $$
 T_F : f\longmapsto \int _VF(v)f(v)dv\quad (f\in {\ss}(V))
 $$ 
 is a tempered distribution. If $T$ is a tempered distribution and there is a measurable function $F$ of polynomial growth such that $T=T_F,$ we identify $T$ with $F$ and write $T(v)$ for $F(v)$; $F$ is determined almost everywhere by $T$, everywhere if it is continuous. 
 \medskip
Fourier transforms $f\longmapsto \widehat f={\ff}f$  are defined using the {\it self-dual measures\/} $d_0x, d_0\xi$ which are $(2\pi)^{-\dim(V)}$ times $dx, d\xi,$ where $d$ refers to the standard Lebesgue measure on $V$ obtained by identifying $V$ with ${\r}^n$ via an orthonormal basis for $V$. Thus, for $f\in {\ss}(V)$, 
$$
 \eqalign{({\ff}f)(\xi)&=\widehat f(\xi)=\int_Ve^{-i(x, \xi)}f(x)d_0x\quad (\xi\in V)\cr
f(x)&=\int_V({\ff}f)(\xi)e^{i(x,\xi)}d_0\xi\quad (x\in V).\cr}
 $$
Given a tempered distribution $T,$ its Fourier transform $\widehat T={\ff}T$ is defined by
$$
\langle {\ff}T, f\rangle=\langle \widehat T, f\rangle\buildrel \rm def \over =\langle T, \widehat f\rangle=\langle T, {\ff}f\rangle.
$$
\medskip
Given a real polynomial $p$ on $V$, the function $F=e^{ip}$ is bounded  and smooth, hence $T_F$ is a tempered distribution. We may therefore consider its Fourier transform $\widehat T_F=\widehat {e^{ip}}$. It is now  natural to ask if $\widehat T_F$ is a function.
\bigskip\noindent
{\bf Definition. \/}{\it
 A real polynomial $p$ on $V$ has the {\it Airy property\/} if 
\medskip\item {1)} there is a smooth function $A_p$ (necessarily unique) of moderate growth on $V$ such that $\widehat {e^{ip}}=T_{A_p}$
\smallskip\item {2)} $A_p$ extends to an entire function on $V_{\c}$, the complexification of $V$.}
\bigskip\noindent
In terms of $e^{ip},$ this is equivalent to saying that
$$
\int _Ve^{ip(y)}({\ff}f)(y)d_0y=\int_VA_p(x)f(x)d_0x\quad (f\in {\ss}(V)).
$$
This can be written out as 
$$
\int _Ve^{ip(y)}d_0y\int _Vf(x)e^{-i(y,x)}d_0x= \int_VA_p(x)f(x)d_0x
$$
which is a rigorous formulation of the formal relation
$$
\int _Ve^{ip(y)-i(x,y)}d_0y=A_p(x).
$$  
The classical polynomial $p(y)=y^3/3$ has the Airy property in this sense, as does the matrix Airy function.
\bigskip\noindent
{\bf 3. Sufficient conditions for a polynomial to have the Airy property.\/} Our first concern is to obtain sufficient conditions for a real polynomial $p(y_1,y_2,\dots ,y_n)$ on ${\r}^n$ to have the Airy property. Note that if $p$ is linear or depends only on a proper subset of the variables $y_1,y_2, \dots ,y_n$, the Airy integral is a distribution on ${\r}^n$ that is supported by a proper affine subspace,  hence is not a function. If $p$ is quadratic, its leading term can be diagonalised so that $e^{ip}$ is a product of functions $e^{ic_jy_j+id_jy_j^2}$, thus reducing the problem to the one-dimensional case. The fact that $cy+dy^2$ has the Airy property when $d\not=0$ is verified by direct evaluation of suitable gaussian integrals. Therefore $p$ has the Airy property when $p$ is of degree $2$ and its leading term is a non-degenerate quadratic form. We thus only consider polynomials of degree at least $3$. When $n=1,$ any non-zero polynomial of degree at least $3$ has the Airy property. In higher dimensions, we must assume some suitable invariance properties to establish the Airy property. Before proving the higher-dimensional versions, we briefly discuss two examples in the one-dimensional case: the classical example when $p(y)=y^3/3,$ and the example when $p(y)=y^4/4$; they  require different treatments. The proof in the higher-dimensional case has a similar structure to that of the one-dimensional case. 
\bigskip\noindent
{\bf 3.1. Case of polynomials on ${\r}$.\/} 
Here $V={\r}$, and $p(y)$ is a real polynomial of degree at least $3$.
\medskip
{\bf The classical case $p(y)={y^3\over 3}$.\/} Let $E=e^{iy^3/3},$ which we view as a tempered distribution. Then we have the differential equation $E'=iy^2E$, which becomes $A^{\prime \prime }+xA=0$ under the Fourier transform,  where $A=\widehat E$. Any tempered solution of this differential equation is a multiple of $\widehat E$; in fact, if $S$ is such a solution, then $S=\widehat a,$ where $a'=iy^2a,$ and so the distribution $a$ is a classical smooth function and is a multiple of $A$. The equation $A^{\prime\prime}+xA=0$ makes sense in the complex plane, is linear, and has no singularities in the finite plane;  hence $A$ is entire. Note that the space of solutions to $A^{\prime\prime}+xA=0$ is two-dimensional, but the tempered solutions form a one-dimensional subspace.
\medskip
To see that $A$ is an entire function of moderate growth on ${\r},$ we change the path of integration in the formal definition of $A$ to the path $C_t$ that goes from $-\infty +it$ to $\infty +it$, where $t>0$. Let 
$$
A(t, z)=\oint _{C_t}\psi(\zeta, z)d\zeta=\int _{-\infty }^\infty \psi(\xi+it, z)d\xi
$$
where
$$
\psi (\zeta, z)=\exp \{(i/3)\zeta^3-iz\zeta\}\quad (\zeta,z\in {\c}, \zeta=\xi+i\eta, z=x+iy).
$$ 
Then
$$
|\psi (\zeta, z)|=\exp \{ -{\rm Im} \left ((1/3)\zeta^3-z\zeta\right )\}=\exp \{-\eta \xi^2+(1/3)\eta^3+y\xi+x\eta\}.
$$
For fixed $L>0$, it follows that $\psi$ and its derivatives with respect to $z$ and $\eta$ are majorized on the path $C_t$ by
$$
K_L (1+|\xi|)^n\exp \{-t \xi^2+L|\xi|\},
$$
valid for $|z|\le L$, for all $\xi\in {\r}$, for all $\eta$ with $0<\eta\le L$, and for   a suitable integer $n\geq 0$. Therefore the integral defining $A(t, z)$ converges absolutely and uniformly for $|z|\le L$ and defines a holomorphic function of $z$ for any $t >0$;  its derivatives with respect to $z$ and $t$ can be calculated by differentiating under the integral sign. 
\medskip
To prove that $A$ does not depend on $t,$ we differentiate with respect to $t$. Thus
$$
{\partial A(t, z)\over \partial t}=\int _{-\infty}^\infty {\partial \psi (\xi+it)\over \partial t}d\xi=i\int_{-\infty}^\infty {\partial \psi (\xi+it)\over \partial \xi}d\xi=0, 
$$
since $\psi(\xi+it)$ vanishes at $\xi=\pm \infty$. So we write
$$
A(z)=\int _{-\infty }^{\infty}\psi(\xi+it, z)d\zeta.
$$
Then
$$
{d^nA\over dx^n}(x)=\int _{-\infty}^\infty(-(\xi+it))^n\psi (\xi+it, z)dt \qquad (n\geq 0).
$$
To see that this derivative has polynomial growth for $z$ on the real axis, we exploit the fact that $t$ can be chosen to {\it depend on $x$\/}.  Let $t=1/|x|$. A simple calculation shows that for $x\ge 1,$ there are constants $K$ and $K'$ such that
$$
\bigg|{d^nA\over dx^n}(x)\bigg|\le K\int_{-\infty}^\infty (1+|\xi|)^ne^{-{\xi^2\over x}}d\xi=K'x^{{n+1\over 2}}.
$$
\medskip
It remains to show that $A$ is the Airy function. We show that
$$
\int _{-\infty}^\infty A(x)f(x)dx=\int_{-\infty}^\infty e^{i{y^3\over 3}}\widehat f(y)dy\qquad (f\in {\ss}({\r})).
$$
 Since $A$ has been shown to have polynomial growth, it defines a tempered distribution, so  it is enough to show this for $f\in C_c^\infty ({\r})$. Let $f\in C_c^\infty ((-a,a))$, where $a>0$. Then $\widehat f$ is defined over ${\c}$, is entire, and is given by
$$
\widehat f(\zeta)=\int_a^af(x)e^{-i\zeta x}dx\quad (\zeta=\xi+i\eta).
$$
We have
$$
|\widehat f(\zeta)|\le e^{a\eta}||f||_1
$$
and the Paley-Wiener estimate 
$$
|\widehat f(\zeta)|\le C_{r,b}(1+\xi^2)^{-r}\quad (r\ge 0, 0<\eta\le b),
$$
where $C_{r,b}>0$ is a constant.
For any $t>0,$
$$
\int_{-\infty}^\infty A(x)f(x)dx=\int_{-\infty}^\infty  f(x)dx\int_{-\infty}^\infty e^{i{(\xi+it)^3\over 3}-ix(\xi+it)}d\xi.
$$
The integrand on the left is majorized by
$$
C_t|f(x)|e^{|x|t}e^{-\eta\xi^2}
$$
(where $C_t>0$ is a constant) which is integrable with respect to $dxd\xi$ and so 
$$
\int_{-\infty}^\infty A(x)f(x)dx=\int_{-\infty}^\infty e^{i{(\xi+it)^3\over 3}}\widehat f(\xi+it)d\xi.
$$
Since this is true for any $t>0,$ we take the limit, as $t$ tends to $0,$ under the integral sign. To justify this, we cannot use the exponential majorant $e^{-t\xi^2},$ on which we have relied so much, because $t$ is now tending to $0$. Instead we use the Paley-Wiener estimate on $\widehat f$ to estimate the integrand.  Since, for $0<t<1,$ the integrand on the right has the majorant
$$
C(1+\xi^2)^{-r},
$$
we can take the limit under the integral sign, giving 
$$
\int_{-\infty}^\infty A(x)f(x)dx=\int_{-\infty}^\infty e^{i{\xi^3\over 3}}\widehat f(\xi)d\xi,
$$
which is what we want.
\bigskip
{\bf The case $p(y)={y^4\over 4}$.\/} The above method fails when we work with ${y^4\over 4}$ because the main term in the majorant of the exponential is $e^{-t \xi^3},$ which is not integrable on $(-\infty, 0)$. {\it In order to carry this argument forward, the path of integration must be changed so that the exponent stays negative.\/} We define
$$
\psi(\zeta, z)=\exp \{(i/4)\zeta^4-iz\zeta\}\quad (\zeta, z\in {\c}, \zeta=\xi+i\eta, z=x+iy).
$$
The path  of integration, $D_t$, is ${\rm Im}(\zeta)=-t$ for $\xi$ from $-\infty$ to $\xi=-k$ for some $k>0$, then an arbitrary path from $-k-it$ to $h+it,$ where $h>0$, and then the path ${\rm Im}(\zeta)=t$ from $\xi=h$ to $\xi=\infty$. The integral does not depend on the choice of $h,\,k,$ or the auxiliary path between $-k-it$ to $h+it$. So, to simplify matters, we set $h=k=1$ and take the auxiliary path to be linear. Thus $D_t$ will be the path 
$$
D_t : \xi\to \delta(\xi)=\xi+ia(\xi)t\qquad a(x)=\cases {{\rm sgn}(x) & if $|x|> 1$\cr x & if $|x|\le 1$.\cr}
$$
Here ${\rm sgn}(x)=x/|x|$.

\epsfxsize=5cm
\centerline{\epsffile{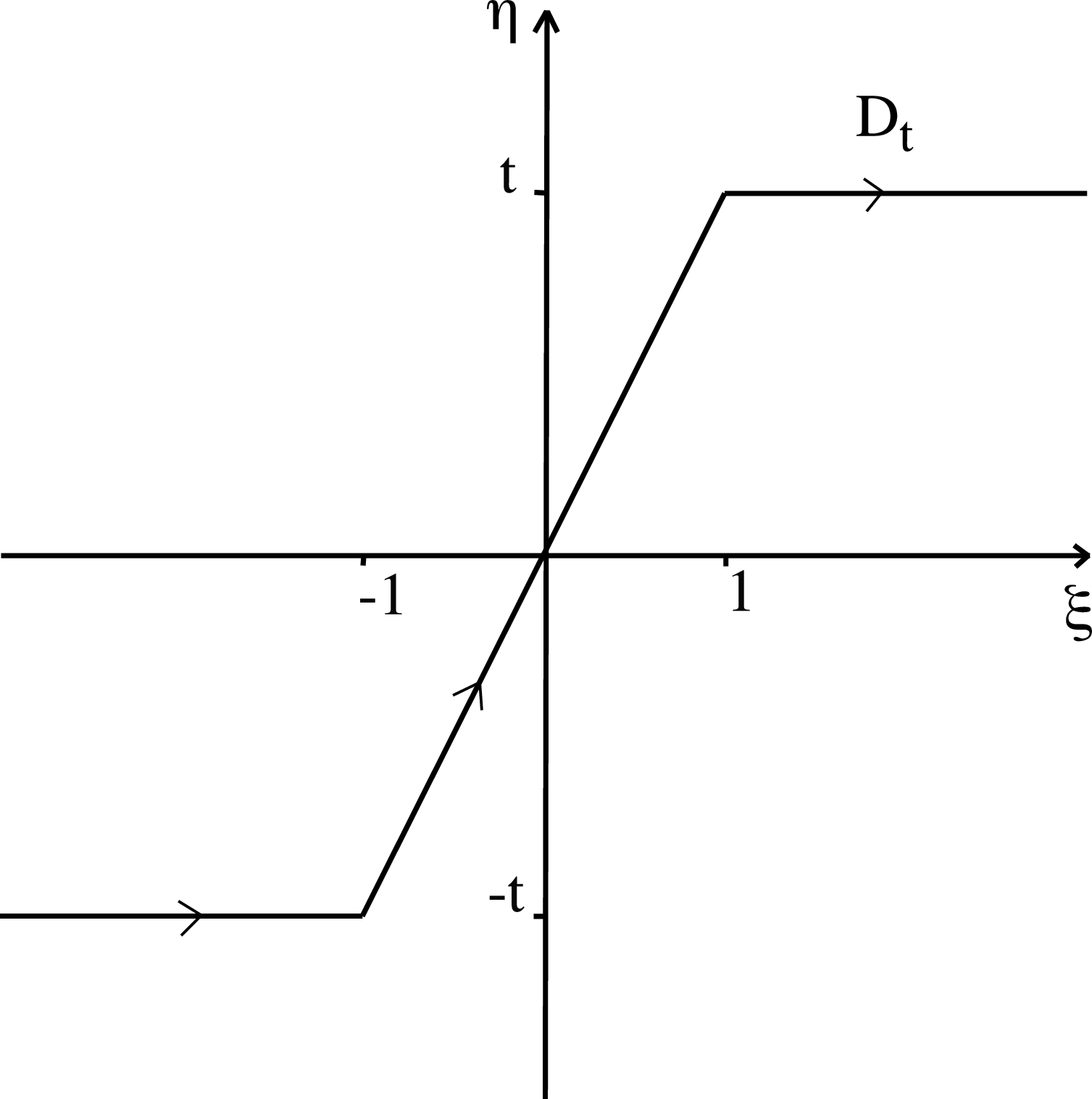}}
\medskip
Define
$$
A(t, z)=\oint _{D_t} \psi (\zeta, z)d\zeta.
$$
To establish convergence of the integral, we note the estimate
$$
|\psi (\xi+i\,{\rm sgn}(\xi)t, z)|\le K_Le^{-t|\xi|^3+(L+1)|\xi|}\eqno (\ast)
$$
valid for $|\xi|\ge 1, |z|\le L, 0<t\le 1$. It follows that if $t_1$ and $t_2$ are fixed with $0<t_1<t_2<1$, then for any $N\ge 0$,
$$
|\psi (\xi+i\,{\rm sgn}(\xi)t, z)|=O(|\xi|^{-N})\qquad (|\xi|\to \infty)\eqno (\ast\ast)
$$
uniformly for $t_1\le t\le t_2$
\medskip
To prove that this integral does not depend on $t,$ we note the relation
$$
D_{t_1}(-A, B)=D_{t_2}(-A, B)+V_{-A}+V_B,
$$
where $D_t(-A, B)$ is the part of the path $D_t$ from $-A$ to $B$ and $V_{-A}, V_B$ are the suitably oriented vertical segments connecting $-A-it_1, -A-it_2$ and $B+it_1, B+it_2$. So the difference between the integrals over the partial paths is at most the sums of absolute values of the integrals over $V_{-A}$ and $V_B$. These tend to $0$ as $A, B\to \infty$ because of the uniform estimate $(\ast\ast)$. The rest of the proof is essentially the same as the proof in the case of degree $3$. 
\medskip
The above arguments apply to the case of an arbitrary real polynomial of degree at least three.
\bigskip\noindent
{\bf 3.2. General case of polynomials on ${\r}^n\, (n\ge 1).$\/} The discussion above can be carried over to show that the results for $y^3/3$ and $y^4/4$ are true for any real polynomial $p(y)$ of degree at least $3$. But more can be done.  Recall that a real homogeneous polynomial is {\it elliptic} if it vanishes only at the origin. We write $S_n$ for the unit sphere of ${\r}^n$ and $S_n^+$ for the subset of $S_n$ where all the coordinates are non-negative. 
\bigskip\noindent
{\bf Theorem 3.2.1.\/} {\it Let $p$ be a real polynomial of degree $m\ge 3$ in    $n$ variables $y_1, y_2, \dots , y_n,$ and let $p_m$ denote its homogeneous component of degree $m$.   Then $p$ has the Airy property in the following cases:

$($1$)$ If $m$ is odd and there is a direction $\tau\in {\r}^n$ such that $\tau{\cdot}\nabla p_m$ is elliptic and strictly positive on $S_n$.

$($2$)$ If $m$ is even, $p_m$ is elliptic and  invariant under arbitrary sign changes of the variables $y_j$, and all the $\partial p_m/\partial y_j (1\le j\le n)$ are non-negative on $S_n^+$. 

Moreover, for any integer $r\ge 0$, all derivatives of $A_p$ of order $r$ are $O(|x|^{(r+n)/(m-1)})$ as $|x|\to \infty$.\/}
\bigskip\noindent
{\bf Remark.\/} It is easy to see that $p$ has the Airy property if and only if $-p$ has the Airy property. Indeed, if $f\mapsto f^\ast$ is the involution defined by $f^\ast(x)=f(-x)^{\rm conj}$, then $A_{-p}=A_p^\ast$. So $p$ has the Airy property if either $p$ or $-p$ satisfies the conditions of the theorem. It is also easy to see that if we partition the set of variables $(y)$ as $(Y, Z)$ and if  $p(y)=p_1(Y)+p_2(Z)$ for some polynomials $p_1$ and $p_2$, then $p$ has  the Airy property if the $p_i (i=1,2)$ have  it; moreover, $A_p=A_{p_1}\otimes A_{p_2}$. Thus $\sum _i c_i y_i^m$ has the Airy property for arbitrary non-zero choices of the constants $c_i$. However no such trick seems to be available to settle the case for $y_1^2y_2^2$. It is an interesting question whether the Airy property is {\it generic\/}; we have not been able to answer it, though we do prove that the property is valid on some non-empty open set (Corollary 3.2.4). 
\bigskip
The next corollary shows that there are plenty of polynomials satisfying the hypotheses of the theorem.
\bigskip\noindent
{\bf Corollary 3.2.2.\/} {\it If $m$ is even, $m=2k$, let $p_m=\sum_{|\alpha|=k}c_\alpha y^{2\alpha},$ where the coefficients $c_\alpha$ are non-negative and the coefficients of $y_j^{2k}$ are strictly positive for $1\le j\le n$.

If $m$ is odd, $m=2k+1$, let $p_m=\sum_{|\alpha|=2k+1}c_\alpha y^{\alpha},$ where all the coefficients $c_\alpha$ are non-negative, and the coefficients of $y_j^{2k+1}$ are strictly positive for $1\le j\le n$. 

Then $p=p_m+q,$ where $q$ is arbitrary and $\deg (q)<m,$ has the Airy property. In particular, $p=y_1^m+\dots +y_n^m+q$ has the Airy property if $\deg(q)<m$.}
\bigskip\noindent
{\bf Proof.\/} The conditions are verified trivially when $m$ is even. When $m$ is odd, take $\tau=(1,1,\dots ,1)$.  $\qed$
\bigskip\noindent
{\bf Corollary 3.2.3.\/} {\it Let $p_m$ be as in Theorem 3.2.1. If $q_1, \dots ,q_r$ are homogeneous polynomials of degree $m,$ and $q$ a polynomial of degree $<m,$ then there exists $\varepsilon >0$ such that 
$$
p=p_m+\sum _{1\le i\le r}c_iq_i+q
$$
has the Airy property for all $c_i\in {\c}$ with $|c_i|<\varepsilon$ for all $i$.\/}
\bigskip\noindent
{\bf Remark.\/} This follows from the proof of the theorem and  is given in \S 3.4. 
\bigskip\noindent
{\bf Corollary 3.2.4.\/} {\it The set of polynomials of degree $N$ that have the Airy property contains a non-empty {\it open} subset of the space of polynomials of degree at most $N$.\/}
\bigskip\noindent
{\bf Proof.\/} This follows immediately from Corollary 3.2.3.
\bigskip\noindent
{\bf 3.2.3. Structure of the proof of Theorem 3.2.1.\/} Since the proof has many technical aspects, it may be worthwhile to sketch the main steps before we undertake the details.
\medskip
{\it Step 1.\/} Since $e^{ip(\xi)-i(\xi, x)}$ is not integrable on ${\r}^n,$ we push ${\r}^n$ into an open cycle\footnote{$^1$}{\mysmall The term {\eightit cycle} usually refers to a map of a compact manifold into
${\c}^n$;  here we shall use the term {\eightit open cycle}  (or {\eightit integration cycle}) to refer to a map of ${\r}^n$ into ${\c}^n$.  This is a
generalisation of the usual notion of a contour for the case $n=1$.} in ${\c }^n$ and transfer the integration to that cycle. The nature of the cycle depends on the parity of the degree of $p$. On this cycle, $e^{ip}$ has a super-gaussian majorant of the form $e^{-c|\xi|^{m-1}}$,  for some constant $c>0$, where $m$ is the degree of $p$. The first step is to define these cycles and prove the majoration.
\medskip 
{\it Step 2.\/} We construct a one-parameter family of cycles $D(t)\,(0<t<1)$ such that $D(t)\to {\r}^n$ as $t\to 0+,$ with $e^{ip}$ having a super-gaussian majorant on the cycles $D(t)$. We define
$$
A(t,z)=\oint_{D(t)}e^{i\left (p(\zeta) -\zeta{\cdot}z\right )}d\zeta\qquad (t>0, \,z,\zeta\in {\c}^n).
$$
We show that $A(t,z)$ is entire in $z$. Then we prove that $A(t,z)$ is independent of $t$ so that it can be written  as $A(z)$. 
\medskip {\it Step 3.\/} Here we prove that $A(x)\, (x\in {\r}^n)$ is of moderate growth. We estimate $A(x)$ and its derivatives by writing it as $A(t,x)$ and {\it choosing $t$ to depend on $x$\/}; for instance, for $|x|>1, t=1/|x|$. 
\medskip
{\it Step 4.\/} By Step 3, $A$ defines a tempered distribution on ${\r}^n$. We now prove that this tempered distribution is $\widehat {e^{ip}}$. Since both distributions are tempered, it is sufficient to prove that they agree for all $f$ that are {\it compactly supported\/}, i.e., we  prove that
$$
\int _{{\r}^n}A(t,x)f(x)dx=\int_{{\r}^n}e^{ip(\xi)}\widehat f(\xi)d\xi.
$$
Since $f$ has compact support, its Fourier transform $\widehat f$ is entire and satisfies the Paley-Wiener estimates. So we take the integration to ${\c}^n$ over the cycles $D(t)$ and let $t\to 0+$. Since $t$ is tending to zero, the super-gaussian majorants can no longer be used; we instead use the Paley-Wiener estimates on $\widehat f$ to carry out the limit.
\bigskip\noindent 
{\bf 3.3. Proof of theorem when $p$ is of odd degree.\/} Let $z=x+iy,  \zeta=\xi+i\eta \in {\c}^n$. Since
$$
|e^{ip(\xi+i\eta)}|=e^{-{\rm Im}\, p(\xi+i\eta)},
$$
finding majorants for $|e^{ip}|$ is the same as finding lower bounds for ${\rm Im}\, p$.
\bigskip\noindent
{\bf Lemma 3.3.1 (Super-gaussian majorant).\/} {\it Let $p$ be a real polynomial of odd degree $2k+1\,(k\geq 1)$ and $p_{2k+1}$ be its homogeneous component of degree $2k+1$.  Suppose that for some $\tau\in {\r}^n$, $\tau{\cdot}\nabla p_{2k+1}$ is elliptic and strictly positive on $S_n$ (the unit sphere of ${\r}^n$).  Then there exist constants $b, C>0$ such that for all $t \in (0,1)$ and for all $\xi\in{\r}^{n}$ with $|\xi|\ge 1$, 
$$ 
{\rm Im}\,p(\xi+it\tau)\ge Ct\big (|\xi|^{2k}-b|\xi|^{2k-1}\big ).
$$
In particular, there exist constants $D, K>0$ such that
$$
|e^{ip(\xi+it\tau)}|\le Ke^{-Dt|\xi|^{2k}}
$$
uniformly for all $\xi\in {\r}^n, 0<t<1$.\/}
\bigskip\noindent
{\bf Proof.\/} In this and the next subsection we use the Taylor expansion of a real polynomial $f$, 
$$
f(\xi+i\eta)=\sum _\alpha {(i\eta)^\alpha\over \alpha!}f^{(\alpha)}(\xi)
$$
where $\alpha$ runs over multi-indices and the notation, imitating what happens in the case of one variable, is standard. In particular, for real $f$ and  $\xi, \eta \in {\r}^n$,
$$
{\rm Im \,}f(\xi+i\eta)=\sum _{|\alpha | {\rm odd}}(-1)^{(1/2)(|\alpha|-1)}{\eta^{\alpha}\over \alpha !} f^{(\alpha)}(\xi).
$$
\medskip
Write $p=p_{2k+1}+q$ where $\deg(q)\leq 2k$. Then there exists a constant $C$ such that
$$
\tau \cdot \nabla p_{2k+1}(\xi)\ge C>0
$$
on the unit sphere $S_n$, and as $\tau\cdot\nabla p_{2k+1}$ is homogeneous of degree $2k$, 
$$
\tau\cdot\nabla p_{2k+1}(\xi) \geq C|\xi|^{2k}\qquad (\xi \in {\r}^n).
$$
Since $\deg(q)\le 2k$, there exists  $a>0$ such that uniformly for $t\in (0, 1)$, $|\xi| \ge 1$, 
$$
|{\rm Im} \,q(\xi+it\tau) |\leq ta|\xi|^{2k-1}.
$$
Hence, for $t\in (0, 1), |\xi|\ge 1$, 
$$
{\rm Im} \,p(\xi+it\tau)\ge  {\rm Im} \,p_{2k+1}(\xi+it\tau)-ta|\xi|^{2k-1}
$$
while
$$
{\rm Im}\,p_{2k+1}(\xi+it\tau)= t\bigg((\tau\cdot\nabla) p_{2k+1}(\xi)+O(|\xi|^{2k-2})\bigg) .
$$
Hence there is a constant $b>0$ such that for $t\in (0, 1), |\xi|\ge 1$, 
$$
{\rm Im} \,p(\xi+it\tau)\ge Ct\big(|\xi|^{2k}-b|\xi|^{2k-1}\big). 
$$

The uniform estimate for $|e^{ip}|$ follows immediately by a standard argument.  $\qed$ 
\medskip
We now prove Theorem 3.2.1 for $p$ as above. Write
$$
\psi_p(\zeta, z)=e^{ i\left (p(\zeta)-\zeta{\cdot}z\right)}.
$$
Then
$$
|\psi_p(\xi+it\tau)|=e^{t\tau{\cdot}x+\xi{\cdot}y}|e^{ip(\xi+it\tau)}|\eqno (\ast \ast).
$$
We want to push the integration cycle from ${\r}^n$ (where the integral does not exist) to the cycle ${\r}^n+it\tau$ where the integral converges because of Lemma 3.3.1. We thus define ($C_t$ is defined at the beginning of \S3.1)
$$
A_{p}(t,z)=\oint _{C_t\times \dots \times C_t} \psi_p(\zeta, z) d\zeta_1\dots d\zeta_n=\int_{{\r}^n}\psi(\xi+it\tau)d\xi.
$$
From Lemma 3.3.1 and $(\ast \ast)$ we see that for some constant $K_1=K_1(L)$ we have, uniformly for $|z|\le L, \xi\in {\r}^n, t\in (0,1)$,
$$
|\psi _p(\xi+it\tau, z)|\le K_1e^{-ct|\xi|^{2k}+L|\xi|}.
$$
Therefore the integral defining $A_p$ is absolutely convergent, uniformly when $z$ is bounded, and so defines an entire function of $z$. The derivatives of $\psi_p$ with respect to $z$ and $t$ have similar majorants, except that there will be additional polynomial growth terms of the form $(1+|\xi|)^r$ that do not affect convergence. Hence the derivatives of $A_p$ can be calculated by differentiating under the integral sign. 
\medskip
We now prove that {\it the function $A_{p}(t,z)$ is independent of $t$;\/}  we show that 
$$
{\partial \over \partial t}A_p(t,z)=0.
$$
Now,
$$
{\partial \over \partial t}A_p(t,z)=\int _{{\r}^n}{\partial \psi _p(\xi+it\tau,z)\over \partial t}d\xi=\int _{{\r}^n}\sum _{1\le j\le n}i\tau_j
{\partial\psi_p(\xi+it\tau, z)\over \partial \xi_j}d\xi.
$$
Since $\psi_p$ vanishes at infinity, 
$$
\int_{{\r}^n} {\partial\psi_p(\xi+it\tau, z)\over \partial \xi_j}d\xi=0.
$$
This completes the proof. $\qed$
\medskip
We thus write
$$
A_p(z)=\int _{{\r}^n}\psi_p(\xi+it\tau, z)d\xi.
$$
For the next step in the proof of the theorem we must show that $A_p(x)$ and all of its derivatives have polynomial growth for $z=x\in {\r}^n$. If we write $\partial^{(r)}$ for $\partial_1^{r_1}\dots \partial_n^{r_n}$ where $\partial_j=\partial/\partial z_j$ and $(-i\zeta)^{r}$ for $(-i\zeta_1)^{r_1}\dots (-i\zeta_n)^{r_n}$, then 
$$
\partial^{(r)} A_{p}(x)
=\int _{{\r}^n}(-i\zeta)^{r}\psi_p(\xi+it\tau, x)d\xi.
$$
Let $|x|>1$ and $t=1/|x|.$ The exponential factor $e^{t\tau{\cdot}x}$ in the bound for $|e^{ip}|$ is now $e^{\tau{\cdot}x/|x|}$ and so is bounded by $e^{|\tau|}$. Hence, by Lemma 3.3.1, with $|r|=r_1+r_2+\dots +r_n$, there exists $K_1$ such that
$$
|{(-i\zeta)^{r}\psi_{p}(\xi+i|x|^{-1}\tau,x)|\le K_1(1+|\xi|)^{|r|} e^{-c|x|^{-1}|\xi|^{2k}}.
}
$$
We now make the transformation $\xi\to c^{1/2k}|x|^{-1/2k}\xi$ to get
$$
|\partial^{(r)} A_{p}(x)|\le K_2|x|^{(|r|+n)/2k}\int _{{\r}^n}(1+|\xi|)^{|r|}e^{-|\xi|^{2k}}d\xi
$$
for some constant $K_2$,
which shows that
$$
\partial^{(r)} A_{p}(x)=O(|x|^{(|r|+n)/2k})\qquad (|x|\to \infty).
$$
This proves that $A_p$ is a function of moderate growth and so defines a tempered distribution. 
\medskip
It remains to show that this tempered distribution is the Fourier transform of $e^{ip}$. This is the same as showing that 
$$
\int A_p(x)f(x)dx=\int e^{ip(y)}\widehat f(y)dy.
$$
Since both sides define tempered distributions, it is sufficient to prove this equality for functions $f$ that are compactly supported.  The idea is to work with $A_p(t,x)$ and let $t\rightarrow 0$.  Since $t$ becomes arbitrarily small in the proof, we cannot use the super-gaussian majorants;  we rely instead on the Paley-Wiener estimates for $\widehat f,$ which are available {\it since $f$ is compactly supported.\/}  
\medskip
Assume that ${\rm supp}(f)\subseteq [-\alpha,\alpha]^{n}$ for some $\alpha>0$. Then,
$$
\widehat f(\zeta)=\int \limits_{{\r}^n} f(x)e^{-ix\cdot \zeta}dx = \int \limits_{[-\alpha,\alpha]^n} f(x)e^{-ix\cdot \zeta}dx.
$$
Hence $\widehat f$ is well-defined for all $\zeta\in {\c}^n$ and entire. From Lemma 3.3.1 and $(\ast \ast)$ we have, for any $t>0,\, \xi\in {\r}^n,\, x\in [-\alpha, \alpha]^n$, 
$$
|\psi_p(\xi+it\tau, x)|\le K_2e^{-ct|\xi|^{2k}}.
$$
Hence $f(x)\psi_{p}(\xi+it\tau,x)$ is integrable with respect to $dxd\xi$.  So, for any $t > 0$,
$$
\eqalign{
\int \limits_{\r^n} A_{p}(x)f(x)dx  
&= \int \limits_{[-\alpha, \alpha]^n} A_{p}(t,x)f(x)dx  \cr
&= \int \limits_{[-\alpha, \alpha]^n} \bigg( \int \limits_{\r^n} \psi_{p}(\xi+it\tau,x) d\xi \bigg)f(x)dx  \cr
&= \int \limits_{{\r}^n} e^{ip(\xi+it\tau)} \widehat{f}(\xi+it\tau)d\xi  
}
$$
by Fubini's theorem. We wish to take the limit $t\rightarrow 0$. Since $t$ is tending to $0$ from above, we may assume that $0 < t <1$.  It follows from Lemma 3.3.1 that
$$
|e^{ip(\xi+it\tau)}|\le K_3.
$$
By the Paley-Wiener estimate, for any $r \ge 1$ and uniformly for $0 < t \le 1$,
$$
|\widehat{f}(\xi+it\tau)| \leq C_{r}(1+|\xi|^{2})^{-r}.
$$
Therefore, 
$$
| e^{ip(\xi+it\tau)} \widehat{f}(\xi+it\tau)| \le K_4(1+|\xi|^{2})^{-r}.
$$
This estimate justifies taking the limit under the integral sign. The theorem is thus fully proved when the degree of $p$ is odd and $p_m$ satisfies the appropriate condition. 
\bigskip\noindent
{\bf 3.4. Proof of  theorem when $p$ is of even degree.\/} Here the integration cycles are more complicated than in the odd case, and so the  derivation of the lower bounds for ${\rm Im }\, p$ becomes more involved. Let 
$$
T^+=\{{\theta} =(t_1, \dots , t_n)\in {\r}^n\ \big |\ 0<t_j<1\ {\rm for\ all}\  j\}
$$
and 
$$
 \max ({\theta})=\max_jt_j, \qquad \min({\theta})=\min_jt_j.
 $$ 
Let 
 $$
 a(x)=\cases {{\rm sgn}(x) & if $|x|> 1$\cr x & if $|x|\le 1$\cr}\qquad 
 a(\xi){\theta}=(a(\xi_1)t_1, \dots ,a(\xi_n)t_n).
$$ 
For $t_1>0, \dots ,t_n>0,$ let $D _{\theta}:{\r}^n\longmapsto {\c}^n$ be the map defined by
$$
D_{\theta}(\xi)=\xi+ia(\xi){\theta}.
$$
Then $D_{\theta}$ is an open cycle in ${\c}^n,$ written  $D_{\theta}=D_{t_1}\times \dots \times D_{t_n}$. On this cycle,
$$
d\zeta_j=(1+it\chi(\xi_j)t_j)\,d\xi_j\buildrel \rm def \over =b(\xi_j, t)\,d\xi_j,
$$ 
where $\chi(\xi)$ is the characteristic function of $(-1, 1)$ and $|b(\xi_j, t)|\le 2$; moreover $|\xi|\le |\zeta|\le |\xi|+n$.
\medskip
Suppose that $f$ is an entire function on ${\c}^n$ such that for all $N\ge 0$,
$$
f(\zeta)=O(|\zeta|^{-N})=O(|\xi|^{-N})\qquad (N>n, \,\zeta\in D_{\theta}, \,|\zeta|\to \infty)\eqno (\dag \dag)
$$
for each $\theta$. The integral 
$$
\oint _{D_{\theta}}f(\zeta) d\zeta=\int_{{\r}^n}f(D_{\theta}(\xi))b(\xi_1,t_1)\dots b(\xi_n, t_n)d\xi
$$
is then absolutely convergent.
\bigskip\noindent
{\bf Lemma 3.4.1.\/} {\it Let $f$ satisfy $(\dag \dag)$ uniformly when $\theta$ varies over compact subsets of $T^+$. Then 
$$
\oint _{D_{\theta}}f(\zeta_1, \dots ,\zeta_n) d\zeta_1\dots d\zeta_n
$$
does not depend on  $\theta$.}
\bigskip\noindent
{\bf Proof.\/} Write $I(t_1, t_2,\dots ,t_n)$ for the integral. Let $(s_1,\dots , s_n)\in T^+$. We must prove that $I(t_1,\dots ,t_n)=I(s_1,\dots ,s_n)$. By the absolute convergence we can evaluate these as repeated integrals. If we fix $t_1,\dots ,t_{n-1}$, then  $\{(t_1,\dots ,t_{n-1}, t)\ |\ t\in [s_n,t_n]\}\subset T^+$ is a compact set, and so, $f$ as a function of $\zeta_n$ (when the other $\zeta_j$ are fixed) is $O(|\zeta_n|^{-N})$ uniformly for $t\in [s_n, t_n]$. Hence we can conclude as in \S 3.1 that 
$$
\oint_{D_{t_n}}f(\zeta)d\zeta_n=\oint_{D_{s_n}}f(\zeta)d\zeta_{n}.
$$
We then do the same by integrating with respect to $\zeta_{n-1}$ to change $t_{n-1}$ to $s_{n-1}$, and so on. $\qed$
\medskip
Let $N=\{1,2,\dots ,n\}.$ For any subset $I$ of $N$, let $\xi_I=(\xi)_{i\in I}.$ We can identify ${\r}^n$ with ${\r}^{|I|}\times {\r}^{|N\setminus I|}$ via the map $\xi\mapsto (\xi_I, \xi_{N\setminus I})$.  The following lemma is fundamental in deriving lower bounds for Im($p(\xi+ia(\xi)\theta)$ in Lemma 3.4.3 below.
\bigskip\noindent
{\bf Lemma 3.4.2.\/} {\it Let $p_{2k}$ be a homogeneous real polynomial of degree $2k$ such that $p>0$ and all the derivatives $\partial_jp_{2k}=\partial p_{2k}/\partial \xi_j \,(j=1,2,\dots ,n)$ are non-negative on $S_n^+$. Then there exist $\gamma, c>0$ with the following property. If $I\subset N$ and $I\not=N$, then 
$$
\sum _{j\in N\setminus I} \partial_j p_{2k}(\xi_I, \xi_{N\setminus I})\ge c |\xi|^{2k-1}\qquad (\xi_j\ge 0, |\xi|\ge 1, |\xi_I|\le \gamma ).
$$
}
\bigskip\noindent
{\bf Proof.\/} It is sufficient to prove the estimate for each fixed $I$. Let $S(\xi):=S(\xi_I, \xi_{N\setminus I})=\sum _{j\in N\setminus I}\partial_j p_{2k}(\xi_I,\xi_{N\setminus I})$. We claim that $S(0,\xi_{N\setminus I})>0$ for all $(0, \xi_{N\setminus I})\in S_n^+$.  If this is not the case, then all the partial derivatives involved, being $\ge 0$, will be $0$ at some $(0,\xi_{N\setminus I}^0),$ so that, by Euler's theorem on homogeneous functions, $p_{2k}(0,\xi_{N\setminus I}^0)=0$, contradicting the assumption on $p_{2k}$. Therefore there exists $c>0$ such that $S(0,\eta)\ge 2c>0$ for $(0, \eta)\in S_n^+$, and, by continuity, for some $\gamma >0$,  $S(\xi_I, \xi_{N\setminus I})\ge c$ for all $\xi\, ( \xi_j\ge 0)$ with $|\xi_I|\le \gamma,\, 1-\gamma\le |\xi_{N\setminus I}|\le 1+\gamma$. If $|\xi_I|\le \gamma$ and $|\xi|=1$, we have $|\xi_{N\setminus I}|\in [1-\gamma, 1+\gamma],$ so that $S(\xi_I, \xi_{N\setminus I})\ge c>0$ when $\xi_j\ge 0, |\xi_I|\le \gamma$ and $|\xi|=1$. If now $|\xi_I|\le \gamma, |\xi|\ge 1$, and $\xi'=|\xi|^{-1}|\xi$, then $|\xi'|\le \gamma$ and $|\xi'|=1$, so that $S(\xi_I, \xi_{N\setminus I})=|\xi|^{2k-1}S(\xi'_I, \xi'_{N\setminus I})\ge c|\xi|^{2k-1}$. $\qed$
\bigskip\noindent 
{\bf Lemma 3.4.3. (Super-gaussian majorant).\/} {\it Let $p$ be a real polynomial of even degree $2k\,(k\geq 1)$ and  $p_{2k}$ be the homogeneous component of $p$ of degree $2k$.  Suppose that
\medskip\itemitem 
{1.} $p_{2k}$ is elliptic and positive.
\smallskip\itemitem 
{2.} All the derivatives $\partial_j p_{2k}\, (j=1,2,\dots ,n)$ are non-negative on $S_n^+$.
\smallskip\itemitem 
{3.} $p_{2k}$ is  invariant under all sign changes.
\medskip\noindent
Then there exist constants $A>0$ and $d>0$ such that for all ${\theta}\in T^+,$ and for all $\xi\in {\r}^{n}$ with $|\xi|\ge 1$, 
$$ 
{\rm Im}\,p(\xi+ia(\xi){\theta})\geq d\big (\min ({\theta})|\xi|^{2k-1}-A\max({\theta})|\xi|^{2k-2}\big ).
$$
\/}
\bigskip\noindent
{\bf Proof.\/} As in the odd case, for any real polynomial $f$, we have
$$
{\rm Im}\,f(\xi+i\eta)=\sum _{|\alpha | {\rm odd}}(-1)^{(1/2)(|\alpha|-1)}{\eta^{\alpha}\over \alpha !} f^{(\alpha)}(\xi).
$$
Then, if $r$ is the degree of $f$,
$$
|{\rm Im }f(\xi+ia(\xi){\theta})|\le A\max ({\theta})|\xi|^{r-1}
$$
and, for some constant $A>0$,
$$
{\rm Im}\,f(\xi+ia(\xi){\theta})\ge \sum _ja(\xi_j)t_j\partial_jf(\xi)-A\max({\theta})|\xi|^{r-3}
$$
for $|\xi|\ge 1, {\theta}\in T^+$. Write $p=p_{2k}+q$ where $\deg(q)\leq 2k-1.$ 
\medskip
We first derive the lower bound for ${\rm Im}\,p_{2k}(\xi+ia(\xi){\theta})$.  Assume first that all the $\xi_j\ge 0$, $|\xi|\ge 1$. Then, 
$$
{\rm Im}\,p_{2k}(\xi+ia(\xi){\theta}) \geq \sum _ja(\xi_j)t_j \partial _jp_{2k}(\xi)-A\max({\theta})|\xi|^{2k-3}.
$$
Let $c>0, \gamma >0$ be as in Lemma 3.4.2; we may assume $\gamma <1$. If $I$ is the set of all $j$ such that $\xi_j< \gamma /n$, then $I$ is a proper subset of $N$ (otherwise $\xi|\le \gamma <1$), and $|\xi_I|\le \gamma$, and so 
$$
 \sum _ja(\xi_j)t_j \partial _jp_{2k}(\xi)\ge (\gamma /n)\min({\theta})  \sum _{j\in N\setminus I}\partial _jp_{2k}(\xi)\ge c(\gamma/n)|\xi|^{2k-1}.
 $$
Thus, with $d=c(\gamma /n)$,
$$
{\rm Im}\,p_{2k}(\xi+ia(\xi){\theta}) \geq d\min({\theta})|\xi|^{2k-1}-A\max({\theta})|\xi|^{2k-3}
$$
for $|\xi|\ge 1, {\theta}\in T^+$. This is the estimate we want but with all $\xi_j\ge 0$. 
\medskip
If $\xi_j$ are of variable signs, let $\xi_{j_1}, \dots \xi_{j_r}$ be all those that are strictly negative. Then
$$
\eqalign {{\rm Im}\,p_{2k}(\xi+i\,a(\xi){\theta})&={\rm Im}\,p_{2k}(\dots ,\xi_{j_\nu}+ia(\xi_{j_\nu})t_{j_\nu},\dots )\cr
&={\rm Im}\,p_{2k}(\dots ,-\xi_{j_\nu}+ia(-\xi_{j_\nu})t_{j_\nu},\dots )\cr}
$$
where all the real parts are now non-negative and so can be estimated by the preceding result. Hence
$$
{\rm Im}\,p_{2k}(\xi+ia(\xi){\theta}) \geq  d\min({\theta})|\xi|^{2k-1}-A\max ({\theta})|\xi|^{2k-3}
$$
uniformly for $|\xi|\ge 1, {\theta}\in T^+$. Now $p=p_{2k}+q$ where $\deg(q)\le 2k-1$, so that
$$
|{\rm Im }\,q(\xi+ia(\xi){\theta})|\le A\max ({\theta})|\xi|^{2k-2}.
$$
Thus we finally have the asserted uniform lower bound for ${\rm Im}\,p(\xi+ia(\xi){\theta})$. $\qed$
\bigskip\noindent
{\bf Corollary 3.4.4.\/} {\it If $\Omega \subset T^+$ is a compact set, there are constants $K>0$ and $D>0$ such that for all $\xi\in {\r}^n, \,{\theta}\in \Omega,\, 0<\sigma\le 1$,
$$
|e^{ip(\xi+i\sigma a(\xi){\theta})}|\le Ke^{-D\sigma|\xi|^{2k-1}}.
$$}
\bigskip\noindent
{\bf Proof.\/} We note that there are constants $b>0$ and $b'>0$ such that $\min({\theta})\ge b'>0$ and $\max ({\theta})\le b\min ({\theta})$ when ${\theta}\in \omega$. Then, for $0<\sigma\le 1$ and ${\theta}\in \omega,$ we have $\min(\sigma {\theta})\ge \sigma b'$ and $\max (\sigma {\theta})\le b\sigma \min ({\theta})$, so that for some $D_1>0$,
$$
{\rm Im}\, p(\xi+ia(\xi)\sigma {\theta})\ge D_1\sigma  |\xi|^{2k-1}
$$
for $|\xi|\ge C$. The estimate for $|e^{ip}|$ is now deduced, in a standard manner, from the lower bound for ${\rm Im}\, p$. $\qed$
\medskip
Let $\psi_p$ be defined as before, 
$$
\psi_p(\zeta, z)=e^{i\left (p(\zeta)-\zeta{\cdot}z\right )}.
$$
Then
$$
|\psi_p(\xi+ia(\xi){\theta})|\le Ke^{|{\theta}||x|}e^{-c|\xi|^{2k}+L|\xi|}
$$
uniformly for $\xi\in {\r}^n, {\theta}\in \omega, |z|\le L$. We now define
$$
A_p({\theta}, z)=\oint _{D_{\theta}}\psi_p(\zeta, z)d\zeta=\int_{{\r}^n}\psi_p(D_{\theta}(\xi), z)b(\xi, {\theta})d\xi
$$
From the above estimate for $\psi_p$ we see that the integral converges absolutely and uniformly for $z$ bounded, hence defines an entire function on ${\c}^n$. By Lemma 3.4.1 and Corollary 3.4.4, it is independent of ${\theta}$. So we can write 
$$
A_p(z)=A_p({\theta},z)=\int_{{\r}^n}\psi_p(D_{\theta}(\xi), z)b(\xi, {\theta})d\xi,
$$
where
$$
b(\xi, {\theta})=b(\xi_1, t_1)\dots b(\xi_n, t_n).
$$
\medskip
The proof of the moderate growth of $A_p(x)$ is now exactly as in the odd case. The derivatives of $A_p$ with respect to $z$ are given by differentiating under the integral sign, and we choose ${\theta}$ depending on $x$ by $t^0_j=|x|^{-1}$ for $|x|>1$. Then we have
$$
|{\partial^{(r)}\psi_{p}(\xi+ia(\xi){\theta}^0,x)|\le K_1(1+|\xi|)^{|r|} e^{-c|x|^{-1}|\xi|^{2k-1}}.
}
$$
We now make the transformation $\xi\to c^{1/2k-1}|x|^{-1/2k-1}\xi$ to get
$$
|\partial^{(r)} A_{p}(x)|\le K_1|x|^{(|r|+n)/2k-1}\int _{{\r}^n}(1+|\xi|)^{|r|}e^{-|\xi|^{2k-1}}d\xi,
$$
giving
$$
\partial^{(r)} A_{p}(x)=O(|x|^{(|r|+n)/2k-1})\qquad (|x|\to \infty).
$$
This proves that $A_p$ is a function of moderate growth and so defines a tempered distribution. 
\medskip
The proof that $A_p=\widehat {e^{ip}}$ needs no change from the odd case. We have for compactly supported $f$ and all $0<\sigma <1$, 
$$
\int _{{\r}^n}A_p(x)f(x)dx=\int _{{\r}^n}e^{ip(\xi+ia(\xi)\sigma {\theta})}\widehat f(\xi+ia(\xi)\sigma {\theta})b(\xi, \sigma {\theta})d\xi.
$$
Here ${\theta}\in T^+$ is fixed. We now let $\sigma \to 0+$. Since $b(\xi, \sigma {\theta})$ remains bounded when $\sigma\to 0+$ and tends to $1$, we get
$$
\int _{{\r}^n}A_p(x) f(x) dx=\int _{{\r}^n}e^{ip(\xi)}\widehat f(\xi)d\xi.\quad \qed
$$

\bigskip\noindent
{\bf Proof of Corollary 3.3.3.\/} The lower bound for ${\rm Im}\,p_m$ is of the form $c|\xi|^{m-1}$ for some $c>0$. So, if the $|c_i|$ are sufficiently small, a similar lower bound will hold for $p$ (with $c/2$ in place of $c)$. The rest of the proof remains unchanged. $\qed$
\bigskip\noindent
{\bf 4. Reduction of Airy property to a Cartan subalgebra.\/} In this section we  study Airy polynomials on the Lie algebra $\frak g$ of a connected compact Lie group $G$. The main goal is to prove the theorem that if $\frak h$ is a Cartan subalgebra of $\frak g$, then a $G$-invariant polynomial $p$ on $\frak g$ has the Airy property if $p_\frak h$, its restriction  to $\frak h,$ has the Airy property on $\frak h$. This is quite remarkable and depends on the marvellous way in which analysis on $\frak h$ is related to analysis on $\frak g$; the relationship is a consequence of Harish-Chandra's deep work [HC] on differential operators and Fourier transforms on semisimple Lie algebras. Since this work is not easy to read, we summarize it briefly in the appendix, giving enough details so that the reader can refer to the original work with little difficulty.
\bigskip
{\bf Preliminaries.\/} Let $G$ be a connected compact Lie group, $\frak g$ its Lie algebra, and $\frak h$ a Cartan subalgebra of $\frak g$. Fix a positive system of roots of $(\frak g, \frak h)$, and write $\alpha >0$ to mean that $\alpha$ is a member of the positive system. Let $W$ be the Weyl group of $(\frak g, \frak h)$. We work with a $G$-invariant positive-definite scalar product on $\frak g$. Define
$$
\pi=\prod _{\alpha >0}\alpha.
$$
Then $\pi$ is a polynomial function on $\frak h$ that is skew-symmetric with respect to $W$. Given a function $f$ on $\frak g,$ let $f_\frak h$ denote its restriction to $\frak h$. We need two propositions: Proposition 4.1, concerning division by $\pi$, and Proposition 4.4, relating the behavior of invariant functions on $\frak g$ to that of their restrictions to $\frak h$. 
\bigskip\noindent
{\bf Proposition 4.1.\/} {\it Let $f$ be a smooth function on $\frak h$ that is skew-symmetric with respect to $W$. Then $f=\pi g,$ where $g$ is a smooth $W$-invariant function. Morever, if $f$ is of moderate growth, then  $g$ is of moderate growth.\/}
\bigskip\noindent
{\bf Proof.\/} We begin with a lemma on division by a linear function.
\bigskip\noindent
{\bf Lemma 4.2.\/} {\it Let $f$ be a smooth function on ${\r}^n$ that vanishes on the hyperplane $x_1=0$. Then there is a unique smooth function $g$ such that $f=x_1g$. Moreover, if $f$  is of moderate growth, then $g$ is of moderate growth.\/}
\bigskip\noindent
{\bf Proof.\/} Since $g$ is uniquely defined on the dense open set where $x_1\not=0$, the uniqueness of $g$ is clear. For the existence, since $f(0,x_2,\dots ,x_n)=0$, we have, writing $\partial _1=\partial/\partial x_1$, 
$$
\eqalign {f(x_1,x_2,\dots ,x_n)&=\int _0^1{d\over dt}f(tx_1,x_2,\dots ,x_n)dt\cr &=x_1\int _0^1 \partial_1f(tx_1,x_2,\dots ,x_n)dt\cr}
$$
so that
$$
g(x_1,x_2,\dots ,x_n)=\int_0^1 \partial_1f(tx_1,x_2,\dots ,x_n)dt.
$$
Differentiating this formula shows that if $f$ is of moderate growth, then the same is true for $g$. $\qed$
\bigskip\noindent
{\bf Lemma 4.3.\/} {\it Let $L_1, L_2, \dots ,L_q$ be real linear functions on a real finite-dimensional vector space $V,$ no two of which are proportional. Suppose that $f$ is a smooth function on $V$ that vanishes on the hyperplanes $L_i=0,\, i=1,2,\dots ,q$. Then there is a unique smooth function $g$ such that $f=L_1L_2\dots L_qg$; moreover, if $f$ is of moderate growth, then $g$ is of moderate growth.\/}
\bigskip\noindent
{\bf Proof.\/} When $q=1,$ the result is just Lemma 4.2, since we can take a basis such that $V\simeq {\r}^n$ and $L_1\simeq x_1$. We argue by induction on $q$. Suppose that we can write $f=L_1L_2\dots L_rg_r$ for $r<q$ with $g_r$ smooth. Since the $L_i$ are mutually non-proportional, the set of points on the hyperplane 
$L_{r+1}=0$ where $L_1,L_2,\dots ,L_r$ are all non-zero is a dense open set in this hyperplane, and $g_r$ must vanish on it. Hence $g_r$ vanishes on the whole hyperplane $L_{r+1}=0$. So we can write $g_r=L_{r+1}g_{r+1}$ for a unique smooth function $g_{r+1}$, showing that $f=L_1L_2\dots L_{r+1}g_{r+1}$. If $f$ has moderate growth, then we know, by the inductive hypothesis, that $g_r$ has moderate growth; hence, by Lemma 4.2, $g_{r+1}$ has moderate growth. This completes the inductive argument. $\qed$
\bigskip 
We can now complete the proof of Proposition 4.1. Since $f$ is skew-symmetric on $\frak h$ with respect to $W$, $f$ must vanish on the hyperplane where a root $\alpha$ vanishes; indeed, if $s_\alpha$ is the reflection in that root hyperplane, we have $f=-s_\alpha f$, and so $f=-f$ on the hyperplane, which shows that $f=0$ there. No two roots are proportional; hence we are in the situation treated in Lemma 4.3. So we can write $f=\pi g,$ where $g$ is smooth and unique, with $g$ of moderate growth if $f$ is of moderate growth. Away from where $\pi $ is $0$, the skew-symmetry of both $f$ and $\pi$ implies that $g$ is $W$-invariant; hence $g$ is $W$-invariant on $\frak h$. $\qed$
\bigskip\noindent
{\bf Proposition 4.4.\/} {\it Suppose  that $f$ is a smooth $G$-invariant function on $\frak g$. Then $f$ is of moderate growth on $\frak g$ if and only if $f_\frak h$ is of moderate growth on $\frak h$.\/}
\bigskip\noindent
{\bf Proof.\/} We begin with a simple observation: if $F$ is a smooth $G$-invariant function on $\frak g,$ and if $F_\frak h$ has polynomial growth on $\frak h$, then $F$ has polynomial growth on $\frak g$. To see this, note that we can find a constant $C$ and an integer $r\ge 0$ such that $|F(H)|\le C(1+||H||)^r$ for all $H\in \frak h$. If $X\in \frak g$ we can find $x\in G$ such that $X^x=H\in \frak h$. Then
$$
|F(X)|=|F(H)|\le C(1+||H||)^r=C(1+||X||)^r.
$$
This proves the observation. 
\medskip
Let us now consider the framework of Proposition 4.4. The non-trivial part of the theorem is to show that if $f_\frak h$ has moderate growth on $\frak h$, then  $f$ has moderate growth on $\frak g$, since we must consider the derivatives transversal to $\frak h$. Our proof relies on the theorem of G. Schwarz [S] asserting that any $G$-invariant $C^\infty$-function $f$ on $\frak g$ can be written
$$
f=F(p_1,p_2,\dots ,p_\ell),
$$ 
where the $p_i$ are real generators of the ring of invariant polynomials on $\frak g$ (here $\ell=\dim(\frak h)$) and $F$ is $C^\infty$. Let
$$
p : X\longmapsto (p_1(X),p_2(X), \dots , p_\ell(X))
$$
be the map of $\frak g$ into ${\r}^\ell$. It is easily seen from examples that the image need not be dense in ${\r}^\ell,$ unlike in the complex case, so  $F$ is not uniquely determined. Given any function $g$ on ${\r}^\ell,$ we write $g^\sim$ for the pullback $g\circ p$. Certainly $F^\sim$ is uniquely determined; in fact it is $f$ itself.
\medskip
We begin with a simple chain rule formula for repeated differentiation of composite maps. Let $(x_i)_{1\le i\le m}$ be real variables and 
$$
u(x)=v(q_1(x),q_2(x), \dots , q_k(x)),
$$
where $u, v, q_j$ are smooth functions. We indicate derivatives by successive subscripts so that
$$
u_{i_1i_2\dots i_r}={\partial \over \partial x_{i_r}}\dots {\partial \over \partial x_{i_1}}u.
$$
Let
$$
q : x\longmapsto (q_1(x),q_2(x),\dots ,q_k(x)).
$$
Then the formula that we need is the following:
$$
\eqalign {u_{i_1i_2\dots i_r}&=\sum_{j_1j_2\dots j_r}
\{v_{j_1j_2\dots j_r}\circ q\}\ q_{i_1j_1}\dots q_{i_rj_r}\cr
&\qquad\qquad 
+\sum _{p_1,p_2,\dots p_t, t<r}\{v_{p_1\dots p_t}\circ q\}\ w_{p_1p_2\dots p_t}\cr}
$$
where $w_{p_1p_2\dots p_t}$ are elements in the algebra generated by the derivatives of the $q_j$. When $r=1,$ this is the standard chain rule
$$
u_i=\sum _{1\le j\le k}\{v_j\circ q\}\ q_{ji}.
$$
The general rule is proved easily by induction on $r$.
\medskip
We return to the context of $\frak g, \frak h$ and write
$$
f=F(p_1,p_2,\dots ,p_\ell).
$$
The key is the following lemma.
\bigskip\noindent
{\bf Lemma 4.5.\/} {\it If $t_1,t_2,\dots t_\ell$ are the coordinates on ${\r}^\ell$, then
$$
F_{i_1i_2\dots i_\ell}\circ p
$$
is of moderate growth on $\frak h$ for all $i_1i_2\dots i_r$.\/}
\bigskip\noindent
{\bf Proof.\/} Let $y_1,y_2,\dots y_\ell$ be the coordinates on $\frak h$. We need the classical result (see the Appendix) that
$$
\det (p_{ij}(y))=\pi(y),
$$
where $y$ is the point of $\frak h$ with coordinates $(y_i)$ and
$$
\pi =\prod _{\alpha >0} \alpha
$$
where the product runs through all the positive roots. We use induction on $r$.
\medskip
Let $r=0$. Then $f(y)=F(p(y))$ and so is the restriction to $\frak h$ of the given invariant function on $\frak g$. It is of moderate growth by assumption. So $F\circ p$ is of moderate growth in $y$.
\medskip
Now let  $r$ be $\ge 1$ and assume that the result is proved when the number of differentiations is $<r$. We use the general version of the chain rule proved above and obtain
$$
f_{i_1i_2\dots i_r}=\sum _{j_1j_2\dots j_r}\{F_{j_1j_2\dots j_r}\circ p\}\ 
p_{i_1j_1}\dots p_{i_rj_r}+B_{i_1i_2\dots i_r}
$$
where the terms $B_{i_1i_2\dots i_r}$ are sums of terms of the form
$$
\{F_{p_1p_2\dots p_t}\circ p\}\ w_{p_1p_2\dots p_t}
$$
with $t<r$ and the $w_{p_1p_2\dots p_t}$ are polynomials. So $B_{i_1i_2\dots i_r}$ has moderate growth in $y$ by the induction hypothesis. We now write this as
$$
\sum _{j_1j_2\dots j_r}\{F_{j_1j_2\dots j_r}\circ p\}\ 
p_{i_1j_1}\dots p_{i_rj_r}=f_{i_1i_2\dots i_r}-B_{i_1i_2\dots i_r}
$$
and invert this as a system of linear equations for the $F_{j_1j_2\dots j_r}\circ p$. Since 
$$
\det (p_{ij})=\pi
$$
it follows that 
$$
\pi^r\ F_{j_1j_2\dots j_r}\circ p=\sum Q_{j_1j_2\dots j_r,i_1i_2\dots i_r}(f_{i_1i_2\dots i_r}-B_{i_1i_2\dots i_r})
$$
where the $Q$'s are polynomials. Thus 
$$
\pi^r\ F_{j_1j_2\dots j_r}\circ p
$$
is of moderate growth in $y$, hence, by our previous result, 
$$
F_{j_1j_2\dots j_r}\circ p
$$
is of moderate growth. The induction step is completed and this finishes the proof of the lemma.  $\qed$
\bigskip\noindent
{\bf Completion of the proof of Proposition 4.4.\/} Let $y_i,z_j$ be coordinates on $\frak g$. We write $x=(y,z)=(x_k)_{1\le k\le m}$. Since 
$$
F_{j_1j_2\dots j_r}\circ p
$$
are of moderate growth in $y$, they are trivially of polynomial growth on $\frak h$ and so the corresponding invariant functions on $\frak g$ are of polynomial growth on $\frak g$. By the chain rule formula, if we differentiate repeatedly with respect to the $x$ variables, then, with $k_i$ varying from $1$ to $m$, we have
$$
f_{k_1k_2\dots k_r}=\sum \{F_{j_1j_2\dots j_s}\circ p\} Q_{k_1k_2\dots k_r,j_1j_2\dots j_r},
$$
where the $Q$'s are polynomials. The polynomial growth property of the left side is immediate. $\qed$
\medskip
There is an alternative proof that does not use Schwarz's theorem, instead relying only on Harish-Chandra's work. 
\bigskip\noindent
{\bf Lemma 4.6.\/} {\it If $D$ is a $G$-invariant constant coefficient differential operator on $\frak g$, then $Df$ has polynomial growth on $\frak g$.\/}
\bigskip\noindent
{\bf Proof.\/} We use Harish-Chandra's theory (see the Appendix). We can view $D$ as an element of the symmetric algebra $S(\frak g)$ invariant under $G$. We write $D_\frak h$ for the {\it restriction\/} of $D$ to $\frak h$, namely the unique element in $S(\frak h)$ such that $D\equiv D_\frak h$ modulo the ideal in $S(\frak g)$ generated by the orthogonal complement $\frak h^\perp$ of $\frak h$ with respect to the Killing form. Then
$$
\pi(H)(Df)(H)=(D_\frak h(\pi f_\frak h))(H)\quad (H\in \frak h).
$$
Since $f_\frak h$ has moderate growth on $\frak h$, the same is true of $D_\frak h(\pi f_\frak h)$, so that $\pi (Df)_\frak h$ is of moderate growth on $\frak h$. By Lemma 4.3, we  conclude that $(Df)_\frak h$ is of moderate growth on $\frak h$, hence certainly of polynomial growth on $\frak h$. Since $Df$ is $G$-invariant, it is clear that $Df$ is of polynomial growth on $\frak g$. $\qed$
\bigskip\noindent
{\bf Lemma 4.7. \/} {\it Let $\Delta$ be the Laplacian on $\frak g$ with respect to an orthonormal basis of $\frak g$. Then $\Delta^rf$ has polynomial growth on $\frak g$ for any integer $r\ge 0$.\/}
\bigskip\noindent
{\bf Proof.\/} This is immediate from Lemma 4.6 since $\Delta$ is $G$-invariant. $\qed$
\medskip
Proposition 4.4 now follows at once from Proposition 1 of the Appendix.
\medskip
To prove the restriction principle we need a lemma.
\bigskip\noindent
{\bf Lemma 4.8.\/} {\it If $g$ is an entire function on $\frak h_{\c}$ that is skew-symmetric with respect to $W$, then there is a $W$-invariant entire function $f$ on $\frak h_{\c}$ such that $g=\pi f$. If $f$ is a $W$-invariant entire function on $\frak h_{\c},$ then there is a unique  $G$-invariant entire function $F$ on $\frak g_{\c}$ such that $F|_{\frak h_{\c}}=f$. \/}
\bigskip\noindent
{\bf Proof.\/} Here the suffixes denote complexifications. The proof of the first assertion imitates similar ones in Lemmas 4.2 and 4.3. If $u$ is entire on ${\c}^n$ and vanishes on $z_1=0$, its power series expansion cannot contain any term where $z_1$ does not appear;    so $u=z_1v,$ where $v$ is entire. By a change of coordinates we can replace $z_1$ by any non-zero linear function on ${\c}^n$. Suppose that $L_1, \dots ,L_r$ are linear functions on ${\c}^n,$ no two of which are proportional, and that $u$ is an entire function on ${\c}^n$ that vanishes on the hyperplanes $L_i=0, \,1\le i\le r$. We claim that there is an entire function $v$ such that $u=L_1L_2\dots L_rv$; if $v$ exists,  it is unique, since it is unique on the dense open set where none of the $L_i$ vanish. We prove the claim by induction on $r$. The argument is the same as that in Lemma 4.3.
\medskip
For the second part, let $\alpha$ be a root and $s_\alpha$ be the corresponding Weyl reflection. Since $s_\alpha g=-g$, for each root $\alpha$, $g$ vanishes on the hyperplane where $\alpha =0;$ hence, by the preceding remarks, we can write $g=\pi f,$ where $f$ is entire. The dense open set where no root vanishes is $W$-invariant, and both $g$ and $\pi$ are skew-symmetric with respect to $W$ on it, so  $f$ is $W$-invariant on it, hence on $\frak h_{\c}$. This proves the first statement.
\medskip
For the second,  write $f=\sum _{n\ge 0}p_n,$ where $p_n$ is a homogeneous polynomial of degree $n$. Clearly the $p_n$ are $W$-invariant;  therefore, there are unique homogeneous $G$-invariant polynomials $P_n$ on $\frak g_{\c}$ such that $P_n$ restricts to $p_n$ on $\frak h_{\c}$. Since $f$ is entire, for any $C>0,$ there is a constant $K=K_C>0$ such that the coefficients of the monomials in the power series expansion of $f$ (in a set of linear coordinates on $\frak h_{\c}$) satisfy
$$
|c_{n_1,\dots n_\ell}|\le KC^{-(n_1+\dots n_\ell)}.
$$
Hence, for any $d>0,$ we have an estimate, valid for $H\in \frak h_{\c}$ with $||H||\le d$, 
$$
|p_n(H)|\le K_1C^{-n}d^n n^{\ell -1},
$$
for all $n\ge 0$. This gives 
$$
||P_n(X)||\le K_1C^{-n}d^n n^{\ell -1},
$$
for all $n\ge 0$ and $X\in \frak g_{\c}$ with $||X||\le d$. Since $C$ is  arbitrary, we may choose $C>2d$. This shows that the series $\sum _nP_n$ converges uniformly on compact subsets of $\frak g_{\c}$; hence $\sum_nP_n=F$ is entire and $G$-invariant, and $F$ restricts to $f$ on $\frak h_{\c}$.  
\medskip
We are now ready to state and prove the restriction principle.  
\bigskip\noindent
{\bf Theorem 4.9.\/} {\it Let $p$ be a real $G$-invariant polynomial on $\frak{g}$.  Then $p$ has the Airy property on $\frak{g}$ if  the restriction $p_{\frak{h}}$ of $p$ to $\frak h$ has the Airy property on $\frak{h}$.\/}
\bigskip\noindent 
{\bf Theorem 4.10.\/} {\it Let $A_{p}$ and $A_{p_{\frak{h}}}$ be the Airy functions associated to $p$ and $p_{\frak{h}}$ by Theorem 4.9. Then $\pi^{-1}\partial (\pi) A_{p_{\frak{h}}}$ extends to an entire function on $\frak{h}_{\c}$ and 
$$
A_{p}  \bigg |_{\frak{h}}={1\over \pi}\partial(\pi)A_{p_{\frak{h}}}.  
$$
\/}
\bigskip\noindent
These two theorems are proved together.
\bigskip\noindent
{\bf Proof.\/} We note first that $A_{p_\frak h}$ is $W$-invariant, hence $\partial (\pi)A_{p_\frak h}$ is skew-symmetric with respect to $W$, and $\pi^{-1}\partial (\pi)A_{p_\frak h}$ extends to an entire function, say $B_0$, by Lemma 4.8. Then $\pi B_0=\partial (\pi) A_{p_\frak h}$ is of moderate growth on $\frak h$ and so, by Proposition 4.1, $B_0$ is of moderate growth on $\frak h$. Let $B$ be the $G$-invariant entire function on $\frak g_{\c}$ that restricts to $B_0$ on $\frak h_{\c}$.  By Proposition 4.4, $B$ is of moderate growth on $\frak g$. 
\medskip
For $f\in {\ss}(\frak g)$, 
$$
\phi_{f}(H) = \pi(H) \int _{G} f(u^{-1}Hu)\, du \qquad (H \in \frak{h}).
$$
Then, using the Weyl integration formula and the Harish-Chandra Fourier transform formula (see Appendix), we have (using self-dual measures) 
$$
\eqalign { 
 \left \langle \widehat {e^{ip(Y)}},f(Y)\right \rangle_{\frak{g}} 
&= \int _{\frak{g}} e^{ip(Y)} \hat{f}(Y)d_0Y\cr 
&= c_{W}(\frak{g}) \int _{\frak{h}} e^{ip(H)} \phi_{\hat{f}}(H) \pi(H)d_0H \cr 
&=i^{r}  c_{W}(\frak{g})\int _{\frak{h}} e^{ip(H)} \widehat{\phi_{f}}(H) \pi(H)d_0H \cr
&=i^{r} c_{W}(\frak{g}) <e^{ip(H)},\widehat{\phi_{f}}(H)\pi(H)>_{\frak{h}}\cr 
&=(-i)^{r}i^{r} c_{W}(\frak{g}) <e^{ip(H)},\widehat{\partial(\pi)\phi_{f}(H)}>_{\frak{h}}\cr 
&=(-1)^{r}  c_{W}(\frak{g}) <\widehat{e^{ip(H)}},\partial(\pi)\phi_{f}(H)>_{\frak{h}} \cr 
&= (-1)^{r} c_{W}(\frak{g})  \int _{\frak{h}} A_{p_{\frak{h}}}(H)  \partial(\pi) \phi_{f}(H) d_0H \cr 
&=c_{W}(\frak{g}) \int \limits_{\frak{h}} \partial(\pi) (A_{p_{\frak{h}}}(H) )\phi_{f}(H) d_0H \cr 
&= c_{W}(\frak{g}) \int _{\frak{h}}^{}{1\over \pi(H)} \partial(\pi) (A_{p_{\frak{h}}}(H)) \pi(H) \phi_{f}(H) rd_0H\cr
&=\int _{\frak{g}} B(Y)f(Y)  d_0Y\cr 
&=\left \langle  B(Y),f(Y) \right \rangle_{\frak{g}}\cr}
$$
Note that we do not need the exact integration formula proved in the Appendix, but just the fact that it is valid with a non-zero constant $c_W(\frak g)$. By the remark at the beginning of the proof, $B$ exists, is entire on $g_{\c}$, and is of moderate growth on $\frak{g}$.   But then $p$ has the Airy property, $A_{p}=B$, and $\pi A_{p}|_{\frak{h}}=\partial(\pi)A_{p_{\frak h}}$, as required.
\bigskip\noindent
{\bf 5. Explicit formulas.\/} We apply the results of the previous chapters to obtain an explicit formula for $A_{p}$ for specific invariant polynomials on ${\rm Lie}(G),$ where $G={\rm U}(n)$.
\medskip
In this case, $\frak g=(-1)^{1/2}{\hh}(n),$ where ${\hh}(n)$ is the space of $n\times n$ hermitian matrices. We identify $\frak g$ with ${\hh}(n)$ via the map $X\mapsto (-1)^{1/2}X$ that takes the adjoint action to the usual action of ${\rm U}(n)$ on ${\hh}(n): u, X \mapsto uXu^{-1}$.  Since the theory depends only on the $G$-module that is being considered, it does not matter whether we work with $\frak g$ or ${\hh}(n)$. On ${\hh}(n)$ we take
$$
p(X)={\rm tr}(X^{m})\qquad (X\in {\hh}(n)).
$$
The scalar product on ${\hh}(n)$ is $(X,Y)={\rm tr}(XY)$. 
\medskip
Our results in the preceding section lead to the following. 
$$
A_{p}({\rm diag}(y_{1},\dots,y_{n}))\cdot\prod_{k>\ell}(y_{k}-y_{l})=\prod_{k>\ell}(\partial_{k}-\partial_{\ell}) \bigg ( A_{m}(y_{1})\dots  A_{m}(y_{n})\bigg),
$$
where $A_{m}$ is the one-dimensional Airy function for the polynomial $y^{m}$.
\bigskip\noindent
{\bf Lemma.\/} {\it If ${\ab}$ is a commutative algebra with unit and $a_{k} \in {\ab}, \,1\leq k\leq n$, then
$$
\det (a_i^{j-1})= \prod_{k>\ell}(a_{k}-a_{\ell}).
$$}
\bigskip\noindent
{\bf Proof.\/} This is the classical Vandermonde determinant. $\qed$
\bigskip\noindent
Using the lemma, we can replace $\prod_{k>\ell}(\partial _k-\partial_\ell)$ by $\det (\partial_i^{j-1})$. Expanding this determinant, we obtain
$$
A_{p}({\rm diag}(y_{1},\dots,y_{n}))\cdot\prod_{k>\ell}(y_{k}-y_{l})=
\det(A_m^{(j-1)}(y_i)).
$$
Except for the normalization constant, this is Kontsevich's formula; he uses standard Lebesgue measures whereas we use self-dual measures. To obtain his formula, we proceed as follows. 
\medskip
Let $A_p$ be the Airy function defined with respect to standard Lebesgue measure $dx$. By definition of the Airy property we have  
$$
\int A^{\sharp}_{p}(x)f(x) dx=\int \int e^{ip(y)-i(x,y)}f(x)dx dy ,
$$
and
$$
 \int A_{p}(x)f(x) d_0x=\int \int e^{ip(y)-i(x,y)}f(x) d_0x d_0y.
$$
Hence,
$$
\eqalign{ 
\int A^{\sharp}_{p}(x)f(x)dx&=(2\pi)^{n^2} \int \int e^{ip(y)-i(x,y)}f(x)\, d_0xd_0y \cr
&=(2\pi)^{n^2}\int A_{p}(x)f(x)\, d_0x \cr
&=\int ((2\pi)^{n^2/2}A_{p}(x))f(x)\, dx.
\cr}
$$
Therefore,
$$
A^{\sharp}_p = (2\pi)^{n^2/2}A_{p}.
$$
Similarly,
$$
A^{\sharp}_{p_{\frak{h}}} = (2\pi)^{n/2}A_{p_{\frak{h}}}.
$$
Thus
$$
\pi A_{p}^{\sharp}|_{\frak{h}}=(2\pi)^{n(n-1)/2}\partial(\pi)A^{\sharp}_{p_{\frak{h}}}.
$$
Therefore, 
$$
\prod_{k>\ell}(y_{k}-y_{\ell}). A^\sharp_{p, \frak h}=(2\pi)^{n(n-1)/2}\det(A_{m}^{(j-1)}(y_{i})).
$$
This is precisely the formula that Kontsevich obtained in [K] for $m=3$.
\bigskip\noindent
{\bf 6. Appendix. Invariant differential operators and Fourier transforms on the Lie algebra of a compact Lie group (after Harish-Chandra).\/} Harish-Chandra's work [HC] is a deep study of invariant differential operators and integrals on semisimple Lie algebras. In this appendix, we only examine the case when the Lie group is compact; it need not be semisimple (i.e. ${\rm U}(N))$. 
\medskip
{\bf Structure.\/} Let  $G$ be a connected compact Lie group with Lie algebra $\frak g$. If $\frak h$ is a Cartan subalgebra (CSA), then any element of $\frak g$ can be moved by $G$ to an element of $\frak h$. The centralizer of $\frak h$ in $G$ is a maximal torus $T$ of $G$ and is the connected Lie group defined by $\frak h$.
\medskip
The action of the group $G$ on $\frak g$ gives rise to an action on the algebra $P(\frak g)$ of polynomials on $\frak g$. Let $I(\frak g)=P(\frak g)^G$ be the subalgebra of $G$-invariant elements of $P(\frak g)$. If $\frak h$ is a CSA of $\frak g$, the restriction map $p\mapsto p_\frak h$ is an isomorphism of $I$ with the algebra $I(\frak h)=P(\frak h)^W$ of all polynomials on $\frak h$ that are invariant under the Weyl group $W$; this is a famous theorem of Chevalley. The algebra $I(\frak g)$ has homogeneous generators $p_1, \dots ,p_\ell$ which freely generate it, $\ell$ being the dimension of $\frak h$. They are not uniquely determined, but their degrees are. It is a well-known result that, with respect to linear coordinates $(y_i)$ on $\frak h,$ and writing $q_i=(p_i)_\frak h$, 
$$
{\partial (q_1,\dots ,q_\ell)\over \partial (y_1, \dots ,y_\ell)}=c\pi,
$$
where $c$ is a non-zero constant and $\pi$ is the product of roots in a positive system. This can be seen as follows. If $J$ denotes the Jacobian, then $J$ is a homogeneous polynomial and 
$$
\deg(J)=\sum_{1\le i\le \ell}(\deg(p_i)-1)=\hbox { number of reflexions in } W=\deg (\pi)
$$
(see [V], p. 385). Moreover $J\not=0$ since the $p_i$ are algebraically independent. On the other hand, we can interpret $J(H) \,(H\in \frak h)$ as the determinant of the tangent map $df_H$ where
$$
f : \frak h\longrightarrow {\r}^\ell,\qquad f(H')=(p_1(H'), \dots p_\ell(H')).
$$
If  $s$ is in the Weyl group, then $f=f\circ s;$ hence $df_H=df_{sH}\circ ds_H,$ which shows  that $J(H)=J(sH)\det(s)$. In other words, $J$ is skew-symmetric, hence $J$ is divisible by $\pi$;  as $\deg(J)=\deg (\pi),$ we see that $J=c\pi$ for some non-zero constant $c$.
\medskip
{\bf Polynomial differential operators on a real euclidean space.\/} Let $V$ be a real finite-dimensional euclidean space with scalar product $(\ {\cdot}\ )$. Write $P(V)$ for the algebra of complex polynomials on $V$ and $S(V)$ for the complex symmetric algebra over $V$. Then $S(V)$ acts as the algebra of constant coefficient differential operators on functions on $V$: if we write $\partial(v)$ for the directional derivative along $v\in V$, then, for $v_1, \dots ,v_n\in V$, $\partial(v_1\dots v_n)=\partial(v_1)\dots \partial(v_n)$. The scalar product $(\ {\cdot}\ )$ gives a natural isomorphism $V\simeq V^\ast$ that extends to an isomorphism $P(V)\simeq S(V)$. Given $p\in V,$ we write $\partial(p)$ for the corresponding element of $S(V)$ and view it as a differential operator on $V$. This gives a bilinear form on $P(V)$ defined by
$$
(p,q)=(\partial (p)q)(0).
$$
If $P_m(V)$ is the subspace of elements of degree $m$ in $P(V)$, then $P_m(V)\perp P_n(V)$. If  $(x_i)$ are coordinates on $V$ relative to an orthonormal basis of $V$, then $\partial(x_i)=\partial/\partial x_i$ and
$$
(x_1^{k_1}\dots x_n^{k_n}, x_1^{r_1}\dots x_n^{r_n})=k_1!\dots k_n!\,\delta (k_1,r_1)\dots \delta(k_n,r_n).
$$
In particular, $(\ {\cdot}\ )$ is a non-degenerate  positive-definite scalar product on the subspace of {\it real\/} polynomials on $V$.
\medskip
We write $dV$ or $dx$ for the standard Lebesgue measure on $V$ and $d_0V$ or $d_0x$  for the self-dual measure, so that $d_0V=(2\pi)^{-n/2}dV,$ where $n=\dim (V)$. Fourier transforms are defined with respect to $d_0V$: for $f\in {\ss}(V)$, the Schwartz space of $V$, the Fourier transform is an isomorphsim with itself defined by
$$
\widehat f(y)=\int_Ve^{-i(x,y)}f(x)d_0x,\qquad f(x)=\int_V\widehat f(y)e^{i(x,y)}d_0y.
$$
For any polynomial $p$ on $V$ and $f\in {\ss}(V)$, 
$$
\widehat {pf}=\partial (p^\sim)\widehat f\qquad p^\sim (x)=p(ix).
$$
 Gaussian measures are very useful for computing the constants involved in many formulae. For $g(v)=e^{-(1/2)(x,x)}$, we have $\widehat g(y)=g(y)=e^{-(1/2)(y,y)}$. Moreover $\int _V g(v)d_0v=1$.
\medskip
{\bf Majoration of derivatives by powers of a Laplacian.\/} In classical harmonic analysis there is a principle that all derivatives can be majorized by powers of a Laplacian. This is due to the fact that any polynomial is majorized at infinity by an elliptic polynomial of the same degree. Let $\Delta=\partial _1^2+\partial _2^2+\dots +\partial _n^2$ be the Laplacian, $\partial _i=\partial /\partial x_i$. For any multi-index $\alpha =(\alpha_1,\dots ,\alpha_n),$ we write $\partial^\alpha=\partial_1^{\alpha_1}\dots \partial_n^{\alpha_n}, |\alpha|=\alpha_1+\dots+\alpha_n$, and $x^\alpha =x_1^{\alpha_1}\dots x_n^{\alpha_n}$. 
\bigskip\noindent
The idea of the majoration is very simple. We write $f$ as $(1-\Delta)^{-r}(1-\Delta)^rf$ and use the fact that $(1-\Delta)^{-r}$ is a convolution operator by a function $\phi_r$ with very good properties. The equation $f=\phi_r\ast (1-\Delta)^rf$ can then be differentiated to give 
$$
\partial^\alpha f=\partial^\alpha\phi_r\ast(1-\Delta)^rf
$$
when $|\alpha|\le \ell$. The polynomial growth for $\partial^\alpha f$ follows from this equation and the decay properties of $\partial^\alpha \phi_r$.
\medskip
Let $r$ be a fixed integer with $2r>n$ so that $(1+|x|^2)^{-r}\in L^1(dx)$. Moreover we shall always write $\ell=[2r-n-1]$, the largest integer $<2r-n$. We want to invert $(1-\Delta)^r$. Going over to Fourier transforms, $(1-\Delta)^r$ becomes multiplication by $(1+|\xi|^2)^{r},$ and so its inverse is multiplication by $(1+|\xi|^2)^{-r};$ hence $(1-\Delta)^{-r}$ will be $\phi_r\ast$. We define $\phi_r$ as the Fourier transform of $(1+|\xi|^2)^{-r}$:
$$
\phi_r(x)=\phi_r(-x)=\int {e^{-ix{\cdot}\xi}\over (1+|\xi|^2)^r}d\xi.
$$
The symmetry under $x\to -x$ is seen by changing the integration variable from $\xi$ to $-\xi$. Then $\phi_r$ is of Class $C^\ell$, and for any $\alpha$ with $|\alpha|\le \ell$, 
$$
(\partial^\alpha \phi_r)(x)=\int e^{-ix{\cdot}\xi}{(-i\xi)^\alpha\over (1+|\xi|^2)^r}d\xi\quad (|\alpha|\le \ell).
$$
In particular $\partial^\alpha\phi_r$ is rapidly decreasing as all the derivatives of $(-i\xi)^\alpha (1+|\xi|^2)^{-r}$ are $O((1+|\xi|^2)^{-(2r-\ell)})$. The operator $(1-\Delta)^r$ is then a linear topological isomorphism of ${\ss},$ and its inverse is $\phi_r\ast$. Then for any $f\in {\ss}$ and for any $\alpha$ with $|\alpha|\le \ell,$
$$
(\partial^\alpha f)(x)=(\partial^\alpha\phi_r\ast (1-\Delta )^rf)(x)\quad (x\in {\r}^n, |\alpha|\le \ell).
$$ 
This leads easily to the following proposition.
\bigskip\noindent
{\bf Proposition 1.\/} {\it Suppose that $f$ on that ${\r}^n$ is smooth and $\Delta^sf$ is of polynomial growth for $0\le s\le r,$ where $r$ is an integer $>n/2$. Let $\ell=[2r-n-1]$, the largest integer $< 2r-n$. Then $f$ and its partial derivatives $\partial ^\alpha f$ are of polynomial growth for all $\alpha $ with $|\alpha|\le \ell$. In particular, if $\Delta^sf$ is of polynomial growth for all $s=0,1,2,\dots$, then $f$ is of moderate growth.\/}
\medskip
{\bf The case $V=\frak g$.\/} We are interested in the case $V=\frak g,$ where $(\ {\cdot}\ )$ is a $G$-invariant scalar product on $\frak g$. For instance, for $G={\rm U}(N)$, we take $\frak g=(-1)^{1/2}{\hh}(N),$ and $ (X,Y)= -{\rm tr}(XY)$. Choose a Cartan subalgebra $\frak h$ of $\frak g$. It is known that each element of $\frak g$ can be moved inside $\frak h$ by an element of $G$. Choose a positive system of roots and define
$$
\pi=\prod_{\alpha >0}\alpha .
$$
 Then $\pi$ is a polynomial on $\frak g$. Since each root is pure imaginary on $\frak h$, $i^r\pi$ is real on $\frak h$ and hence $(i^r\pi, i^r\pi)=(-1)^r(\pi, \pi)>0$. $\pi$ is skew-symmetric with respect to $W$. 
 \medskip
 The following lemma is needed to evaluate the derivatives of gaussian functions.
 \bigskip\noindent
 {\bf Lemma 2.\/} {\it Let $E(H)=e^{-(1/2)(H,H)}$. Then
 \medskip\itemitem {$(a)$} $\partial(\pi)E=(-1)^r\pi E$
 \smallskip\itemitem {$(b)$} $(\partial (\pi)^2E)(0)=(-1)^r(\pi, \pi)$
 \smallskip\itemitem {$(c)$} $\widehat {\pi E}=(-i)^r\pi E$.
 \medskip}
 \bigskip\noindent
 {\bf Proof.\/} (a) If $\alpha$ is a root and $H_\alpha$ is its image under the isomorphism $\frak h^\ast \simeq \frak h$ induced by $(\ {\cdot}\ )$, then
 $$
 (\partial (H_\alpha)E)(H)=-\alpha(H)E(H)\qquad (H\in \frak h).
 $$
Hence, by repeated application,  we see that
 $$
 \partial (\pi)E=(-1)^r(\pi +q)E,
 $$
 where $q$ is a polynomial of degree $<\deg (\pi)=r$. Since $E$ is invariant and $\partial(\pi)$ is skew-symmetric with respect to the Weyl group $W$, $\partial (\pi)E$, hence $\pi +q$, is skew-symmetric. But every skew-symmetric polynomial is divisible by $\pi$ and so $q=0$.
 \medskip
 (b) By differentiating (a) we see that
 $$
 \partial (\pi)^2E=(-1)^r\partial (\pi)(\pi).E+F,
 $$
 where $F$ is a sum of terms in each of which some root is present. Hence
 $$
 (\partial (\pi)^2E)(0)=(-1)^r\partial (\pi)(\pi)=(-1)^r(\pi, \pi).
 $$
 \medskip
 (c) We have
 $$
 \widehat {\pi E}=\partial (\pi^\sim)\widehat E=i^r\partial(\pi)E=(-i)^r\pi. \qquad\qquad \qed
 $$
  \bigskip\noindent
{\bf Theorem 3 (Weyl-Harish-Chandra integration formula).\/} {\it For all $f\in {\ss}(\frak g)$, 
$$
\int_\frak gf(X)d_0X =(\pi, \pi)^{-1}\int_\frak h\bigg (\int _Gf(H^x)dx\bigg )\pi(H)^2d_0H, 
$$
where $d_0$ denotes self-dual measure, $H^x={\rm Ad}(x)(H)$, and the measure $dx$ is normalised by $\int_Gdx=1$.\/}
\bigskip\noindent
{\bf Proof.\/} If $\frak h'$ is the subset of $\frak h$ where no root vanishes, then the map $\psi : x, H\mapsto H^x$ induces a map of $G/T\times \frak h'$ into $\frak g$. The complement of the range of $\psi$ is a set of measure $0$ in $\frak g,$ while $\psi$  is a covering map that has bijective differential. It is a standard computation that the determinant of $d\psi$ is $|\pi(H)|^2=(-1)^r\pi(H)^2$. Hence, there is a constant $c>0$ such that the integration formula is valid with $(-1)^rc$ as the constant in front of the right side. The gaussian function $f(X)=e^{-(1/2)(X,X)}$ is used to evaluate $c$. Then 
$$
1=(-1)^rc\int_\frak h\pi(H)^2E(H)d_0H.
$$
Since $\widehat {\pi^2E}=\partial (\pi^\sim)^2\widehat E=(-1)^r\partial(\pi)^2E$, it follows that
$$
\int_\frak h\pi^2Ed_0H==\widehat {\pi^2E}(0)=(-1)^r(\partial(\pi)^2E)(0)=
(\pi, \pi),
$$
so that $(-1)^rc=(\pi, \pi)^{-1}$.  $\qed$
\medskip
{\bf Radial components.\/} An element of $\frak g$ is {\it regular\/} if it is conjugate to an element of $\frak h'$, the subset of $\frak h$ where no root vanishes; we write $\frak g'$ for the dense open set of regular points of $\frak g$. We have observed that the map $\psi : (x,H)\mapsto H^x$ is a covering map from $G/T\times \frak h'$ onto $\frak g'$. Since $d\psi$ is bijective, it follows that given any invariant differential operator $D$ on $\frak g',$ there is a unique differential operator $\delta'(D)$ on $\frak h'$ such that for all smooth invariant functions $f$ on $\frak g$ (it is even enough to take just invariant polynomials),
$$
(Df)_\frak h=\delta'(D)(f_\frak h),
$$
where the suffix denotes restriction to $\frak h$. In analogy with polar coordinates in euclidean space,  $\delta'(D)$ is called the {\it radial component\/} of $D$. The map $D\mapsto \delta'(D)$ is a homomorphism.
\medskip
For $D=\partial(p),$ where $p\in I(\frak g)$, the algebra of $G$-invariant polynomials on $\frak g$, Harish-Chandra computed its radial component. His result is the following beautiful theorem [HC].
\bigskip\noindent
{\bf Theorem 4.\/} {\it For $p\in I(\frak g)$,
$$
\delta'(\partial(p))=\pi^{-1}\circ \partial(p_\frak h)\pi.
$$}
\medskip
{\bf Orbital integrals.\/} For $X\in \frak g, f\in C^\infty(\frak g)$, define
$$
Mf(X)=\int_Gf(X^x)dx\qquad \bigg (\int_G dx=1\bigg).
$$
It is obvious that $Mf$ is $G$-invariant and that $f\mapsto Mf$ is a $G$-equivariant projection operator from $C^\infty(\frak g)$ onto $I^\infty(\frak g):=C^\infty(\frak g)^G$. Harish-Chandra now defines a map $f\mapsto \phi_f$ from $C^\infty (\frak g)$ to $C^\infty(\frak h)$ by
$$
\phi_f(H)=\pi(H)Mf(H)=\pi(H)\int_Gf(H^x)dx.
$$
It is clear that $\phi_f$ is skew-symmetric with respect to $W$. 
\bigskip\noindent
{\bf Theorem 5 (Limit formula).\/} {\it For $f\in C^\infty(\frak g)$,
$$
(\partial(\pi)\phi_f)(0)=(\pi, \pi)f(0).
$$}
\bigskip\noindent
{\bf Proof.\/} We have
$$
(\partial(\pi)\phi_f)(0)=\big (\partial(\pi)\big (\pi Mf\big )\big )(0)
=(\partial (\pi)(\pi)(0)(Mf)(0)=(\pi, \pi)f(0). \quad \qed
$$
\bigskip\noindent
{\bf Lemma 6.\/} {\it If $f$ is a smooth function on a connected open subset $U\subset \frak h$, $H'\in \frak h'$, and $\partial(q)f=q(H')f$ for all $q\in I=P(\frak h)^W$, then there exist constants $c_s\, (s\in W)$ such that
$$
f(H)=\sum _{s\in W} c_se^{(sH, H')}\qquad (H\in U).
$$}
\bigskip\noindent
{\bf Proof.\/} The elements $sH'$ are all distinct as $H'\in \frak h'$. Hence the exponentials $e^{(H, sH')}$ are linearly independent and satisfy the given differential equations. So the solution space has dimension $\ge |W|$. We now argue that the dimension is $\le |W|$. To establish this, we need the result that $P=P(\frak h)$ is a free finite module over $I$ of rank $=|W|$. Let $u_1, \dots ,u_{|W|}$ be a module basis. Let  $f$ be a solution; since $f$ is an eigenfunction of $\Delta,$ where $\Delta$ is the Laplacian (certainly invariant with respect to $W$ if the Laplacian is defined using an orthonormal basis with respect to which $W$ acts orthogonally), $f$ is analytic. If $(\partial (u_j)f)(H_0)=0$ for all $j$ at some point $H_0\in U$, then the relation $P=\sum _jIu_j$ implies that all derivatives of $f$ vanish at $H_0;$ hence $f=0$ on $U$, by analyticity and connectedness of $U$. So the map $f\mapsto ((\partial (u_j)f)_{1\le i\le |W|}$ of the solution space into ${\c}^{|W|}$ is injective, proving that its dimension is $\le |W|,$ and is therefore spanned by the $e^{(H, sH')}$. $\qed$
\bigskip\noindent
{\bf Theorem 7.\/} {\it For $H, H'\in \frak h_{\c}$,
$$
\pi(H)\pi(H')\int_G e^{(H^x, H')}dx=|W|^{-1}(\pi, \pi)\sum_{s\in W}\varepsilon (s)e^{(sH, H')}.
$$}
\bigskip\noindent
{\bf Proof.\/} Here $(\ {\cdot}\ )$ is extended to a complex bilinear form on $\frak g_{\c}\times \frak g_{\c}$; $\varepsilon (s)=\det(s)$ is the homomorphism $W\longrightarrow \{\pm 1\}$ such that $\varepsilon (s_\alpha)=-1$ for all roots $\alpha$.
\medskip
It is sufficient to prove the theorem for $H, H'\in \frak h',$ as both sides are holomorphic. Let $g_{H'}(X)=e^{(X, H')}$. Then $\partial(p)g_{H'}=p(H')g_{H'}$ for $p\in P(\frak g);$ it follows that for if $p\in I(\frak g)$, then $\partial(p) g_{H'}^x=p(H')g_{H'}^x$ for all $x\in G$. Hence $\partial(p)(Mg_{H'})=p(H')Mg_{H'}$. Using radial components we find that
$$
\partial(p_\frak h)(\pi (Mg_{H'})_\frak h)=p_\frak h(H')(\pi (Mg_{H'})_\frak h).
$$
Since $p\mapsto p_\frak h$ is an isomorphism of $I(\frak g)$ with $P(\frak h)^W,$ we can use the lemma to write 
$$
\pi(H)g_{H'}(H)=\sum_{s\in W}c_se^{(sH, H')}\qquad (H\in \frak h^+),
$$ where $\frak h^+$ is a connected component of $\frak h'$. By analyticity, this holds on all of $\frak h$. Since $G_{H'}$ is $W$-invariant, the sum on the right is skew-symmetric and so $c_s=\varepsilon (s)c_1$, so that
$$
\pi(H)g_{H'}(H)=\gamma \sum _{s\in W}\varepsilon (s)e^{(sH, H')},
$$
where $\gamma$ is a constant possibly depending on $H'$. Since $(\partial(\pi)(\pi g_{H'}))(0)=(\pi, \pi)g_{H'}(0)=(\pi, \pi)$, we get $\gamma ={|W|^{-1}(\pi, \pi)\over \pi(H')}$. This proves the theorem. $\qed$
\medskip
It is not difficult to see, from the formula defining $\phi_f,$ that if $f\in {\ss}(\frak g),$ then $\phi_f\in {\ss}(\frak h);$ and that the map $f\mapsto \phi_f$ is continuous from ${\ss}(\frak g)$ to ${\ss}(\frak h)$. Moreover, we can write the integration formula as
$$
\int _\frak g f(X)d_0X=(\pi, \pi)^{-1}\int _\frak h\pi(H)\phi_f(H)d_0H.
$$
\bigskip\noindent
{\bf Theorem 8.\/} {\it Let Fourier transforms on both $\frak g$ and $\frak h$ be defined with respect to self-dual measures. Then
$$
\phi_{\widehat f}=i^r\widehat {\phi_f}.
$$}
\bigskip\noindent
{\bf Proof.\/} By the previous theorem,
$$
\pi(H)\pi(H')(Mg_{H'})(H)=|W|^{-1}(\pi, \pi) \sum _{s\in W}\varepsilon (s)e^{(sH, H')}.
$$
Now, for $f\in {\ss}(\frak g)$, 
$$
\eqalign {\phi_f(H)&=\pi(H)\int_Gf(H^y)dy=\pi(H)\int_G\bigg (\int_\frak g\widehat f(Y)e^{i(Y, H^y)}d_0Y\bigg )dy\cr 
&=\pi(H)\int_\frak g\bigg (\int_Ge^{i(Y, H^y)}dy\bigg )\widehat f(Y)d_0Y\cr
&=\pi(H)\int_\frak g\bigg (\int_Ge^{i(Y^y, H)}dy\bigg )\widehat f(Y)d_0Y\cr
&=\pi(H)\int_\frak g\widehat f(Y)(Mg_{iH})(Y)d_0Y.\cr}
$$
Since $Mg_{iH}$ is bounded, the integrand is in the Schwartz space;  hence we use the integration formula to get
$$
\phi_f(H)=(\pi, \pi)^{-1}\int_\frak h\phi_{\widehat f}(H_1)\pi(H)\pi(H_1)(Mg_{iH})(H_1)d_0H_1.
$$
On the other hand,
$$
\eqalign {\pi(iH)\pi(H_1)(Mg_{iH})(H_1)&=\pi(iH)\pi(H_1)\int_Ge^{i(H^x,H_1)}dx\cr
&=|W|^{-1}(\pi, \pi)\sum _{s\in W}\varepsilon(s)e^{i(sH, H_1)}.\cr}
$$
Therefore,
$$
\eqalign {i^r\phi_f(H)&=|W|^{-1}\int_\frak h\bigg (\sum_{s\in W}\varepsilon (s)e^{i(sH, H_1)}\bigg )\phi_{\widehat f}(H_1)d_0H_1\cr
&= \int_\frak h\phi_{\widehat f}(H_1)e^{i(H, H_1)}d_0H_1.\cr}
$$
This means that
$$
\phi_{\widehat f}=i^r\widehat {\phi_f}.\qquad\qquad \qed
$$
\bigskip\noindent
{\bf Remark.\/} Harish-Chandra's formula in [HC] does not have the correct sign.
\vskip 1 true in
\centerline { \bf References}
\vskip 0.5 true in
\item {[A]} G.B. Airy, {\it On the intensity of light in the neighbourhood of a caustic\/}, Trans. Camb. Phil. Soc. 6 (1838), pp. 379--403. 
\medskip \item {[F]} Rahul N. Fernandez, {\it Airy Functions associated to Compact Lie}

{\it Groups and their Analytic Properties\/}, Ph.D. Thesis (2006), UCLA.
\medskip\item {[HC]}Harish-Chandra, {\it Differential operators on a semisimple Lie algebra\/}, Amer. J. Math 79 (1957), pp. 87--120.
\medskip\item {[K]} Maxim Kontsevich, {\it Intersection theory on the moduli space of curves and the matrix Airy function\/}, Comm. Math. Phys. 147 (1992), no.1, pp. 1--23. 
\medskip\item {[S]} Gerald W. Schwarz, {\it Smooth functions invariant under the action of a compact Lie group\/}, Topology 14 (1975), pp. 63-68. 
\medskip\item {[V]} V.S. Varadarajan, {\it Lie groups, Lie algebras, and their representations\/}, Springer-Verlag, New York, 1984.
\medskip\item {[VS]} Olivier Vallee and Manuel Soares, {\it Airy functions and applications to physics\/} (2004), Imperial College Press, London.
\medskip\item {[W]} Edward Witten, {\it Two-dimensional gravity and intersection theory on moduli space\/}, Surveys in differential geometry (Cambridge, MA), Lehigh Univ., Bethlehem, PA, pp. 243--310 (1991).

\bigskip
\bigskip\noindent
Rahul N. Fernandez, {\it rahul@math.ucla.edu}
\medskip\noindent
V.S. Varadarajan, {\it vsv@math.ucla.edu}

\bye